\documentclass[aps,
prd, 
twocolumn,
nofootinbib,superscriptaddress,floatfix
]{revtex4-1}

\usepackage{natbib}
\usepackage{graphicx}
\pdfoutput=1
\usepackage{subfig,array}
\usepackage{dcolumn}
\usepackage{float}

\usepackage{amsfonts}
\usepackage{amsmath}
\usepackage{overpic}

\renewcommand{\log}{\ln}
\renewcommand{\hat}{\widehat}
\def\M{{\cal M}}
\def\lag{{\cal L}}
\def\Tr{\mathrm{Tr}}
\def\QCD{\text{\tiny QCD}}
\def\SCET{\text{\tiny SCET}}
\def\nnlo{N$^2$LO}
\def\msbar{$\overline{\mathrm{MS}}$}

\newcommand{\ncoll}{$n$-collinear}
\newcommand{\nbcoll}{$\nb$-collinear}

\newcommand{\kAp}{{\vec{k}_{+ t\perp}^{(0)}}{}}
\newcommand{\kA}{ k^{(0)}_{+ t}{}}
\newcommand{\kBp}{{\vec{k}_{- t\perp}^{(0)}}}
\newcommand{\kB}{ k^{(0)}_{- t}{}}
\newcommand{\kAnlo}{k^{(2)}_{+ t}{}}
\newcommand{\kBnlo}{k^{(2)}_{- t}{}}
\newcommand{\kApnlo}{{\vec{k}_{+ t\perp}^{(2)}}{}}

\def\vac#1{\langle0|#1|0\rangle}

\def\nb{{\bar{n}}}
\def\Ln{L_J}
\def\LQ{L_H}
\def\Ls{L_S}
\def\alsb{\bar\al_s}
\def\jqcd{\mathcal{J}_2{}}
\def\mqcd{\mathcal{M}_\QCD{}}
\def\hmqcd{\hat{\mathcal{M}}_\QCD{}}

\def\labelscet{Bauer:2000ew,Bauer:2000yr,Bauer:2001ct,Bauer:2001yt,Bauer:2002nz}

\def\Dslash{\fmslash{D}}
\def\nslash{\fmslash{n}}
\def\nbslash{\fmslash{\bar n}}
\def\wnfun{W_n^{(\bf 3)}}
\def\wnbfun{W_{\bar n}^{(\bf 3)}}
\def\ynfun{Y_n^{(\bf 3)}}
\def\ynbfun{Y_{\bar n}^{(\bf 3)}}
\def\wnadj{W_n^{(\bf 8)}{}}
\def\wnbadj{W_{\bar n}^{(\bf 8)}{}}
\def\ynadj{Y_n^{(\bf 8)}{}}
\def\ynbadj{Y_{\bar n}^{(\bf 8)}{}}
\def\xnb{x_\nb}
\def\xn{x_n}
\def\xnbinf{x_\nb^\infty}
\def\xninf{x_n^\infty}
\def\xsninf{x_{n}^{\infty_s}}
\def\xsnbinf{x_{\nb}^{\infty_s}}
\def\dn{dx^+d^{d-2}x_\perp e^{-iQx^+}}
\def\dnb{dx^-d^{d-2}x_\perp e^{-iQx^-}}

\newcommand{\al}{\alpha}
\newcommand{\ga}{\gamma}
\newcommand{\de}{\delta}
\newcommand{\ep}{\epsilon}

\newcommand{\la}{\lambda}

\newcommand{\si}{\sigma}

\newcommand{\Ga}{\Gamma}

\newcommand{\La}{\Lambda}

\newcommand{\gn}{\mbox{$\gamma_{\nu}$}}

\def\be{\begin{equation}}
\def\ee{\end{equation}}
\def\ba{\begin{eqnarray}}
\def\ea{\end{eqnarray}}

\def\bfig{\begin{figure}[t]}
\def\efig{\end{figure}}
\def\fcaption#1{\caption{\raggedright #1}}
\newcommand{\nn}{\nonumber \\}
\newcommand{\non}{\nonumber}
\def\eqn#1{(\ref{#1})}
\def\fig#1{Fig.\ \ref{fig:#1}}
\def\sec#1{Sec.\ \ref{sec:#1}}
\def\tab#1{Table\ \ref{tab:#1}}
\def\app#1{Appendix\ \ref{app:#1}}
\def\bwtxt{\begin{widetext}}
\def\ewtxt{\end{widetext}}
\def\bop{\begin{overpic}}
\def\eop{\end{overpic}}

\newcounter{ex}
\def\parsec#1{{\bf \stepcounter{ex}\Alph{ex})} $\boldsymbol{ #1}$ :}

\makeatletter
\def\fmslash{\@ifnextchar[{\fmsl@sh}{\fmsl@sh[0mu]}}
\def\fmsl@sh[#1]#2{%
   \mathchoice
     {\@fmsl@sh\displaystyle{#1}{#2}}%
     {\@fmsl@sh\textstyle{#1}{#2}}%
     {\@fmsl@sh\scriptstyle{#1}{#2}}%
     {\@fmsl@sh\scriptscriptstyle{#1}{#2}}}
\def\@fmsl@sh#1#2#3{\m@th\ooalign{$\hfil#1\mkern#2/\hfil$\crcr$#1#3$}}

\makeatother

\begin{document}

\title{Subleading Corrections To Thrust Using Effective Field Theory}
\author{Simon M. Freedman}
\email{sfreedma@physics.utoronto.ca}
\affiliation{Department of Physics, University of Toronto, 
     60 St.\ George Street, Toronto, Ontario, Canada M5S 1A7}
\date{\today}
\begin{abstract}
We calculate the subleading corrections to the thrust rate using Soft-Collinear Effective Theory to factorize the rate and match onto jet and soft operators that describe the degrees of freedom of the relevant scales.  We work in the perturbative regime where all the scales are well above $\Lambda_{\rm QCD}$.  The thrust rate involves an incomplete sum over final states that is enforced by a measurement operator.  Subleading corrections require matching onto not only the higher dimensional dijet operators, but also matching onto subleading measurement operators in the effective theory.  We explicitly show how to factorize the $O(\al_s\tau)$ thrust rate into a hard function multiplied by the convolution of the vacuum expectation value of jet and soft operators.  Our approach can be generalized to other jet shapes and rates.
\end{abstract}

\maketitle


\section{Introduction}\label{sec:intro}

Jet shapes are examples of observables with multiple scales, which can give rise to logarithmic enhancements in the fixed order rate that ruin the perturbative expansion in the strong coupling constant.   Effective Field Theories (EFTs) separate the scales by expanding in their small ratio and matching onto operators that describe the degrees of freedom at each scale.  The logarithmic enhancements  are then summed by using the renormalization group to run between the scales restoring perturbative control of the rate.  The effects of the subleading corrections from the small ratio of scales are systematically calculated in the EFT by matching onto higher dimensional operators.  

The appropriate theory for describing jet shapes with narrow jets is Soft-Collinear Effective Theory (SCET) \cite{\labelscet,Beneke:2002ph, Freedman:2011kj}.  Deriving factorization theorems that separate the scales is straightforward at leading order (LO) in SCET due to the explicit decoupling of the collinear and soft degrees of freedom.  Subleading corrections to SCET have previously been used to study subleading corrections to $B$ decays \cite{Lee:2004ja}; however, jet shapes are complicated by kinematic cuts placed on the phase space.  The goal of this paper is to demonstrate how to factorize the subleading corrections to jet shapes using SCET by considering the example of the thrust observable.

Thrust \cite{Farhi:1977sg} is defined by 
\ba\label{thrust}
	\tau=\frac1Q\sum_{i\in X}\min\left(E_i+\vec t\cdot\vec p_i,E_i-\vec t\cdot \vec p_i\right),
\ea
where $Q$ is the initial energy, $X$ is the final state, and the thrust axis $\vec t$ is the unit vector that maximizes $\sum_{i\in X}|\vec t\cdot\vec p_i|$.  When $\tau\ll1$ the final state is a pair of back-to-back jets with small invariant mass.  Thrust is a convenient observable to illustrate how to calculate subleading jet shapes in SCET due to its simple phase space.  We will concern ourselves only with $e^+e^-\to \gamma^*\to X$ in order to reduce the contribution from initial state radiation and restrict ourselves to vector currents.  It is possible to generalize to weak process but for simplicity, we will not consider axial currents in this paper.  

The LO in $\tau$ thrust rate was calculated using SCET in \cite{Bauer:2008dt,Hornig:2009vb}.   The rate was written in the factorized form
\ba \label{HJS}
	H\times \vac J \otimes\vac{\bar J}\otimes\vac S +O(\tau),
\ea
where the convolution is defined in \sec{lo}.  The LO rate was factorized by matching onto SCET and expanding the final state phase space imposed by \eqn{thrust} in the SCET power counting.  The jet operators, $J$ and $\bar J$, describe the physics at the intermediate scale $\sqrt{\tau}Q$, the soft operator, $S$, describes the physics at the soft scale $\tau Q$, and the hard function, $H$, describes the physics above the hard scale $Q$.  Factorization allows the large $\log\tau$'s to be summed by separately renormalizing the jet and soft operators.  This was done in \cite{Hornig:2009vb}.  

We are interested in extending the results of \cite{Bauer:2008dt, Hornig:2009vb} to include the $O(\tau)$ corrections to the rate.  We restrict our calculation to the perturbative regime when the soft scale is well above $\La_{\rm QCD}$.  We will ignore the effects of hadron masses, which have been discussed in \cite{Mateu:2012nk}.  We will show how the $O(\tau)$ rate can be factorized as in \eqn{HJS} using SCET with the appropriate subleading jet and soft operators.  The se subleading operators are generalizations of the LO operators, and properly separate the scales while having consistent power counting.  We leave renormalizing these operators to a future work.

Understanding how to incorporate subleading phase space effects is important for writing a factorization theorem beyond LO.  We will show how subleading phase space is accounted for in the effective theory by $1)$ consistent expansion of the cuts using the SCET power counting, and $2)$ insertions of subleading operators that account for the QCD momentum conservation expansion.  Both these effects must be accounted for in order to calculate the $O(\tau)$ corrections to the rate and reproduce the perturbative QCD result at $O(\al_s\tau)$ in \cite{Kramer:1986mc}.  

The rest of the paper is organized as follows:  in \sec{scet} we review the description of SCET in \cite{Freedman:2011kj} and explain our reasoning for using this formulation over the formulation in \cite{\labelscet}.   In \sec{lo} we review the LO calculation and introduce the notation used in the rest of the paper.  We calculate the $O(\al_s\tau)$ rate in \sec{nlo} and demonstrate how to write it in a factorized form.  We conclude in \sec{concl}.  The full list of operators is reserved for the appendices. 

For simplicity we use the notation $\LQ\equiv\log(\mu^2/Q^2)$, $\Ln\equiv\log(\mu^2/(Q\sqrt\tau)^2)$, $\Ls\equiv\log(\mu^2/(Q\tau)^2)$, and $\alsb\equiv\al_s C_F/(2\pi)$ where $C_F=(N_C^2-1)/(2N_C)$ for $N_C$ colours.


\section{Review of SCET \label{sec:scet}}

Soft-Collinear Effective Theory describes the interactions between highly boosted and low energy degrees of freedom.  Three types of fields are required for calculating thrust: \ncoll, \nbcoll, and soft, which have characteristic momentum scaling\footnote{The momentum $p^\mu=(p^+,p^-,p_\perp)$ is defined in light-cone coordinates by $p^+\equiv p\cdot n=E-\vec p\cdot\vec n$, $p^-\equiv p\cdot\nb=E+\vec p\cdot\vec n$, and $p_\perp^\mu=p^\mu-p^+\frac{\nb^\mu}{2}-p^-\frac{n^\mu}{2}$ where $n\cdot\nb=2$.} in terms of the SCET expansion parameter $\lambda\ll1$
\ba\label{mmscaling}
	p_n^\mu\sim Q(\lambda^2,1,\lambda), \quad p_\nb^\mu\sim Q(1,\lambda^2,\lambda),\quad  k_s^\mu\sim Q\lambda^2
\ea
respectively.  The two collinear sectors describe fields that are highly energetic and moving in opposite directions, while the soft sector describes low energy fields.  The SCET Lagrangian and operators are derived by expanding the interactions between the different fields in $\lambda$. Momenta denoted by $k$ will be $O(\la^2)Q$ unless otherwise specified.  

\begin{figure}[t]
\centering
	\subfloat[\label{fig:dijetqcd1}]{ \includegraphics[width=0.1\textwidth]{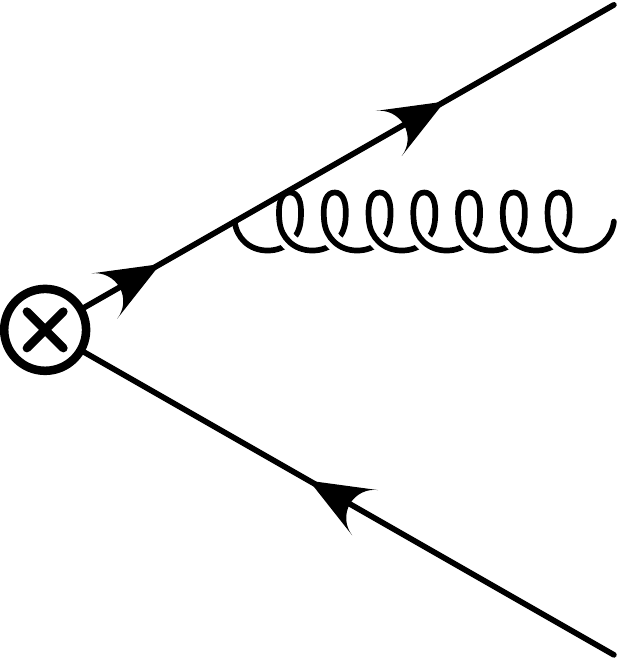}}
	\hspace{0.1\textwidth}
   	\subfloat[\label{fig:dijetqcd2}]{\includegraphics[width=0.1\textwidth]{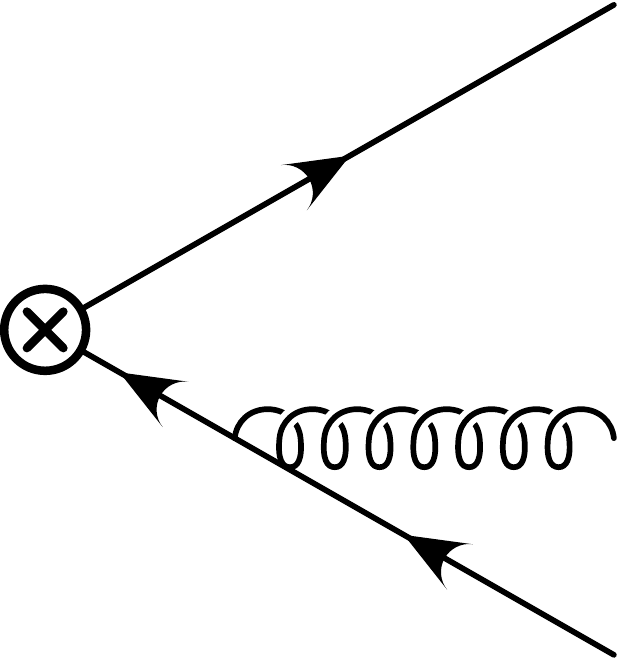}}
   \fcaption{QCD vertex diagrams required for dijet production at $O(\al_s)$. \label{fig:amp}}
\end{figure}

We will use the SCET formulation of \cite{Freedman:2011kj}.  In this formulation, soft and collinear fields are described by QCD in the absence of an external current.  Therefore, the SCET Lagrangian is
\ba\label{SCETlag}
	\lag_\SCET=\lag_\QCD^n+\lag_\QCD^\nb+\lag_\QCD^s,
\ea
which has no subleading contributions and each $\lag^m_\QCD$ will describe fields from the $m$ sector only.  The interactions between the sectors are contained in the external currents. The QCD vector current
\ba\label{jqcd}
	\jqcd^\mu(x)=\bar\psi(x)\gamma^\mu\psi(x)
\ea
is matched onto the SCET dijet operators
\ba\label{qcdmatch}
	\jqcd^\mu\to&&\, C_2^{(0)}O_2^{(0)\mu}+\frac1Q\sum_{i\geq1}C_2^{(i)}O_2^{(i)\mu} +\ldots
\ea
where the superscripts denote the suppression in $\lambda$ and the ellipses represent higher dimensional operators.  The operators and their tree-level matching coefficients $C_2^{(i)}$ are found by expanding the diagrams in \fig{amp} in $\la$ for \ncoll, \nbcoll, and soft fields.  

At LO, the (anti-)quark must be ($\nb$-)\ncoll\ and the gluon can be either collinear or soft.  The LO dijet operator is \cite{Freedman:2011kj}
\ba\label{O2lo}
	O_2^{(0)\mu}(x)&=&\left[\bar\psi_n(\xnb)P_\nb\wnfun(\xnb,\xnbinf)\right]\nn
		&&\times\left[\ynfun(\xsninf, 0)\gamma^\mu\ynbfun(0,\xsnbinf)\right]\\
		&&\times\left[\wnbfun(\xninf,\xn)P_\nb\psi_\nb(\xn)\right],\nonumber
\ea  
with one-loop matching coefficient \cite{Manohar:2003vb}
\ba\label{C2lo}
	C_2^{(0)}(\mu)&=& 1-\alsb\left(\frac12\log^2\left(\frac{\mu^2}{-Q^2}\right)+\frac32\log\left(\frac{\mu^2}{-Q^2}\right)\right.\nn
		&&\left.+4-\frac{\pi^2}{12}\right)+O(\al_s^2)
\ea
and \msbar\ counterterm
\ba
	Z_2^{(0)}(\mu)&=&1+\alsb\left(-\frac1{\ep^2}-\frac3{2\ep}-\frac1\ep\log\left(\frac{\mu^2}{-Q^2}\right)\right)+O(\al_s^2).\nn
\ea
The interactions between the different sectors are reproduced by Wilson lines defined in a representation $\bf R$
\ba\label{wilson}
	W_n^{(\bf R)}(x,y)&=&P\exp\left(-ig\int_0^{\frac n2\cdot(y-x)} ds\,\nb\cdot A_n^a(x+\nb s)T_{\bf R}^a\right)\nn
	Y_n^{(\bf R)}(x,y)&=&P\exp\left(-ig\int_0^{\frac \nb2\cdot(y-x)} ds\,n\cdot A_s^a(x+n s)T_{\bf R}^a\right)\nn
\ea
with similar definitions for $W_\nb^{(\bf R)}$ and $Y_\nb^{(\bf R)}$.  The projectors $P_n=(\nslash\nbslash)/4$ and $P_\nb=(\nbslash\nslash)/4$ are required from the expansion of the Dirac equation.  The positions are defined as
\ba\label{positions}
	\xn=(0,x\cdot\nb,\vec x_\perp)&\quad&\xninf=(0,\infty, \vec x_\perp) \quad \xsninf=(0,\infty,0)\nn
	\xnb=(x\cdot n, 0,\vec x_\perp)&&\xnbinf=(\infty,0,\vec x_\perp) \quad \xsnbinf=(\infty,0,0)\nn
\ea
and are necessary to conserve the appropriate components of momentum that respect \eqn{mmscaling}.  Each square bracket in \eqn{O2lo} is a separately gauge invariant piece, which means the sectors explicitly decouple from one another.   The physical interpretation of the operator is given in \cite{Freedman:2011kj}.

The next-to-leading order (NLO) dijet operators and matching coefficients were found in \cite{Freedman:2011kj} and are reproduced in \app{operators}.  The operators are generalizations of the LO operator \eqn{O2lo} with appropriate derivative insertions.  For $O(\tau)$ corrections to thrust, we will also need the \nnlo\ operators, which are found by following the approach of \cite{Freedman:2011kj}.  These operators and their matching coefficients are also shown in \app{operators}.  Both the NLO and \nnlo\ dijet operators explicitly decouple the sectors in the same way as the LO operator.

A more widely used formulation of SCET \cite{\labelscet} separates the collinear momentum into $p=\tilde p+k$, where $k\sim\lambda^2Q$ and $\tilde p\sim Q,\lambda Q$ are the residual and  label momentum respectively.  The large label momentum is removed from all interactions leading to the Lagrangian being a non-local expansion of two-component spinors with soft and collinear fields explicitly interacting at NLO.  Although, the formalisms of \cite{\labelscet} and \cite{Freedman:2011kj} are equivalent, they approach momentum conservation differently.  This is important when cuts are placed on phase space such as for the thrust rate.  In the approach of \cite{Freedman:2011kj}, momentum is not conserved and the subleading $O_2^{(\de)}$ operators in \app{operators} account for the expansion of QCD momentum conservation using SCET momentum power counting.  In the approach of \cite{\labelscet}, label and residual momentum are separately conserved meaning momentum is exactly conserved.  The equivalent action of the $O_2^{(1\de)}$ operator comes from the subleading kinetic interaction \cite{Bauer:2002uv}
\ba\label{labelnlo}
	\lag_{\xi\xi}^{(1)}\big|_{g_s=0}=\sum_{\tilde p,\tilde p'}\bar\xi_{n,\tilde p}(x)i\fmslash{\partial}^{\perp}\frac{1}{\bar{\mathcal{P}}}\fmslash{\mathcal{P}}^\perp\frac{\nslash}{2}\xi_{n,\tilde p'}(x)+\text{h.c}
\ea
where $\xi_{n,\tilde p}$ is a two-component \ncoll\ spinor with label momentum $\tilde p$, and $\mathcal{P}^\mu$ is an operator that pulls down label momentum.  Insertions of this term into time-ordered products is equivalent to expanding the on-shell condition $(\tilde p+k)^2=0$ in $\la$.  These terms are not necessary for inclusive phase space calculations such as subleading $B$ decays \cite{Lee:2004ja}.  However, they will be important when cuts are placed on phase space such as in the subleading thrust rate and reproduce the action of the $O_2^{(1\de)}$ operator in time-ordered products.  

In this paper we choose to use the formalism of \cite{Freedman:2011kj}.  The Lagrangian of \cite{Freedman:2011kj} has simpler Feynman rules and only one insertion of $O_2^{(1\de)}$ is required instead of an insertion of $\lag_{\xi\xi}^{(1)}$ for each collinear field.  The explicit decoupling of sectors at the operator level also makes it easier to derive a subleading factorization theorem.


\section{Leading Order Calculation \label{sec:lo}}

The LO thrust distribution was calculated using the approach of \cite{\labelscet} in \cite{Hornig:2009vb} and was written in the factorized form \eqn{HJS}.  In this section we review the calculation using the formalism of \cite{Freedman:2011kj}, which gives an equivalent form of the answer.  In the next section we generalize this description to calculate the $O(\tau)$ rate.

The thrust rate is the cumulate of  the distribution
\ba\label{rate}
	R&&(\tau)=\frac1{\si_0}\int d\tau'\frac{d\si}{d\tau'}\theta(\tau-\tau')\\
		&&=\int d^4x e^{-iQ\cdot x} \vac{\jqcd^{\mu\dagger}(x)\hmqcd(\tau)\jqcd_\mu(0)}\nonumber
\ea
where $Q^\mu=(Q/2)(n^\mu+\nb^\mu)$ is the momentum of the incoming photon and $\si_0$ is the Born cross-section.  The measurement operator,  $\hmqcd(\tau)$ \cite{Bauer:2008dt,Bauer:2008jx}, acts on states $|X\rangle$
\ba
	\hmqcd(\tau)|X\rangle\equiv\mqcd(\tau,\{p_X\})|X\rangle
\ea
to project only those final states that give a thrust value $\tau$.  When taking the cuts of diagrams, the function $\mqcd(\tau,\{p_X\})$ generates the appropriate phase space by restricting the momentum of the particles $\{p_X\}$.  

To factorize the rate, we first match the QCD currents and measurement operators onto SCET dijet and measurement operators.  The matching of the QCD currents onto SCET dijet operators was discussed in \sec{scet}.  The QCD measurement operators are matched onto SCET measurement operators in a similar manner
\ba\label{Mmatch}
	\hmqcd(\tau)\to\hat\M^{(0)}(\tau)+\hat\M^{(1)}(\tau)+\hat\M^{(2)}(\tau)+O(\lambda^3),\nn
\ea
where the superscripts refer to the suppression in $\lambda$.  The SCET measurement operators are found by expanding the thrust constraints implemented by $\mqcd$ using the SCET momentum scaling \eqn{mmscaling}.  

Thrust is measured with respect to the thrust axis, $\vec t$, defined below \eqn{thrust}.   The definition of the thrust axis in SCET has an expansion in $\la$ and is written as $\vec t=\vec t^{\,(0)}+O(\lambda^2)$.  The SCET momentum power counting defines the LO thrust axis $\vec{t}^{\,(0)}=-\vec n$ \cite{Bauer:2008dt}, where we have chosen the $-\vec n$ axis to be exactly along the total \nbcoll\ momentum (i.e. $\vec p_{\nb\perp}\equiv 0$).  The overall sign of $\vec t$ is unimportant as seen in \eqn{thrust}.   The sectors decouple in the LO measurement operator \cite{Bauer:2008dt,Hornig:2009vb}
\ba\label{Mfact}
	\hat\M^{(0)}(\tau)=\hat\M_n^{(0)}(\tau_n)\otimes\hat\M_\nb^{(0)}(\tau_\nb)\otimes\hat\M_s^{(0)}(\tau_s)
\ea
because the thrust axis is independent of any individual particle.  The convolution above is defined as
\ba
	&&f_1(\tau_1)\otimes f_2(\tau_2)\otimes f_3(\tau_3)\equiv\\
	&&\int d\tau_1d\tau_2d\tau_3\theta(\tau-\tau_1-\tau_2-\tau_3)f_1(\tau_1) f_2(\tau_2) f_3(\tau_3).\non
\ea
Using the LO definition of the thrust axis, the action of the measurement operators are
\ba\label{Mlocs}
	\M_n^{(0)}(\tau,\{p\})&=&\left(\frac{\sum_i p_i^-}{Q}\right)^{d-2}\delta\left(\tau-\frac1Q\sum_{i}\frac{|\vec p_{i\perp}|^2}{p_i^-}\right)\nn
	\M_\nb^{(0)}(\tau,\{p\})&=&\left(\frac{\sum_i p_i^+}{Q}\right)^{d-2}\delta\left(\tau-\frac1Q\sum_{i}\frac{|\vec p_{i\perp}|^2}{p_i^+}\right)\nn
	\M_s^{(0)}(\tau,\{k\})&=&\delta\left(\tau-\frac1Q\left(\nb\cdot\kA+n\cdot\kB\right)\right)
\ea
where the sums are only over the momentum in each sector.  We have defined
\ba\label{khemi}
	k^{\mu}_{\pm t}=\sum_{i} k_i^\mu\theta(\pm\vec k_i\cdot \vec t\,)=k^{(0)\mu}_{\pm t}+k^{(2)\mu}_{\pm t}+O(\lambda^4)
\ea
as the total soft momentum in the $\pm\vec t$ hemisphere.  The LO definition is
\ba\label{klo}
	k^{(0)\mu}_{\pm t}=\sum_{i} k_i^\mu\theta(\pm\vec k_i\cdot\vec t^{\,(0)})=\sum_i k_i^\mu\theta(\mp\vec k_i\cdot\vec n),
\ea
where in both \eqn{khemi} and \eqn{klo} the sum is over all soft particles.  The $d=4-2\ep$ dependent prefactors in the collinear sectors come from choosing the collinear fields to be in the $n^\mu$ and $\nb^\mu$ directions.  At LO these prefactor have no affect, but are important for the $O(\tau)$ corrections.  

\begin{figure*}[t]
	\centering
	\subfloat[$\vac{J^{(0)}}$\label{fig:j01}]{\bop[width=0.15\textwidth]{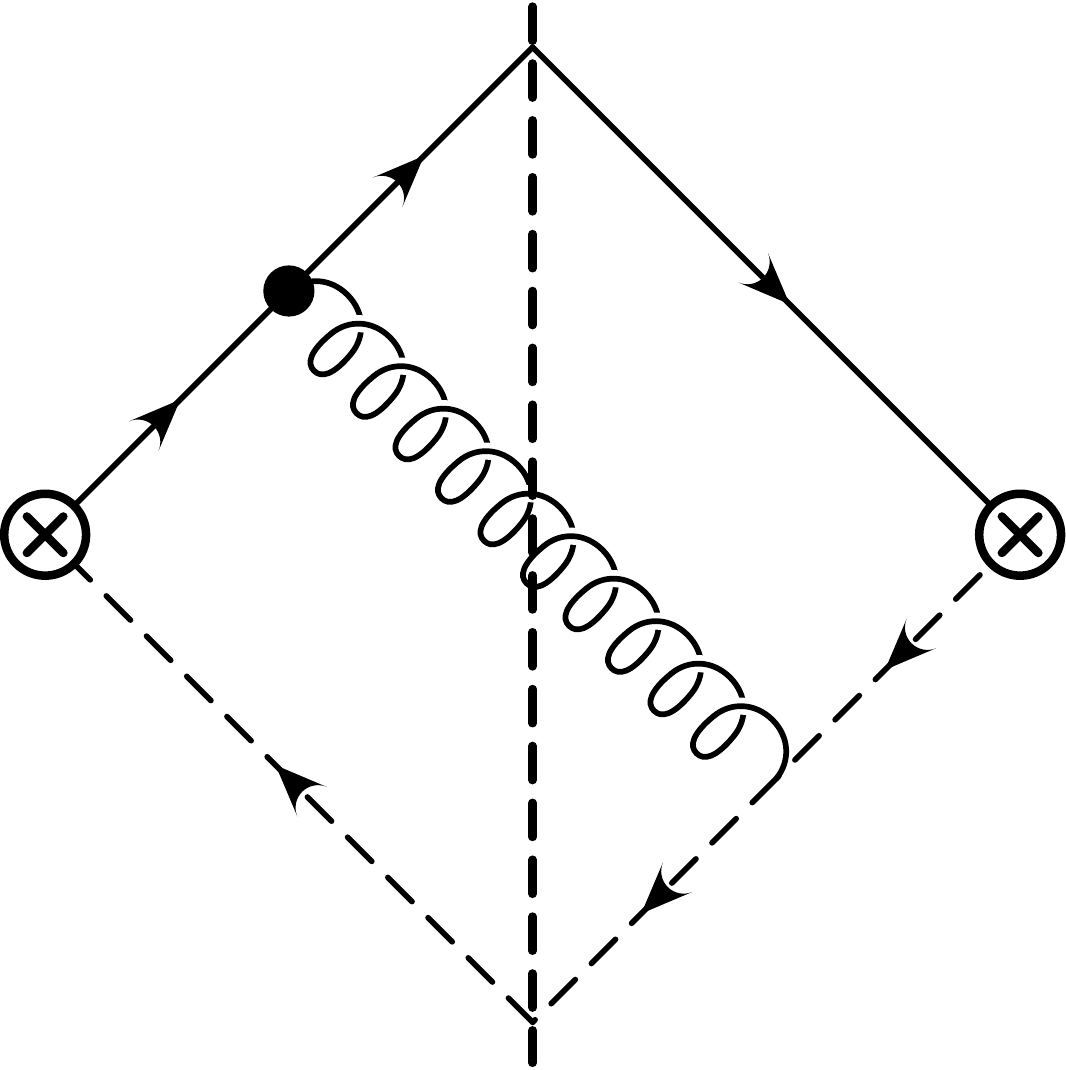}
		\put(0,30){$0$}
		\put(90,30){$x_n$}
	\eop}\hspace{0.05\textwidth}
	\subfloat[$\vac{J^{(0)}}$\label{fig:j01b}]{\includegraphics[width=0.15\textwidth]{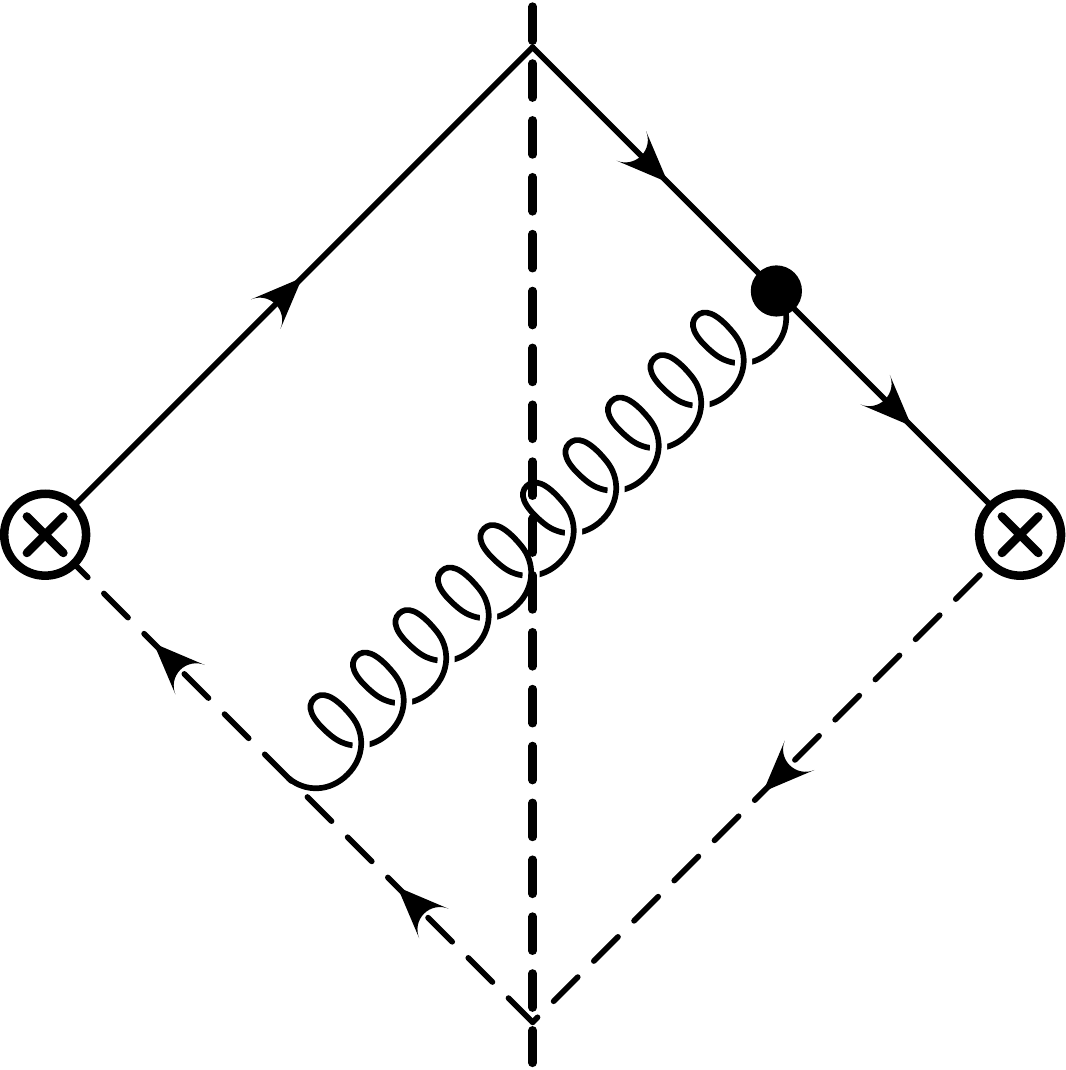}}\hspace{0.05\textwidth}
	\subfloat[$\vac{J^{(0)}}$\label{fig:j02}]{\includegraphics[width=0.15\textwidth]{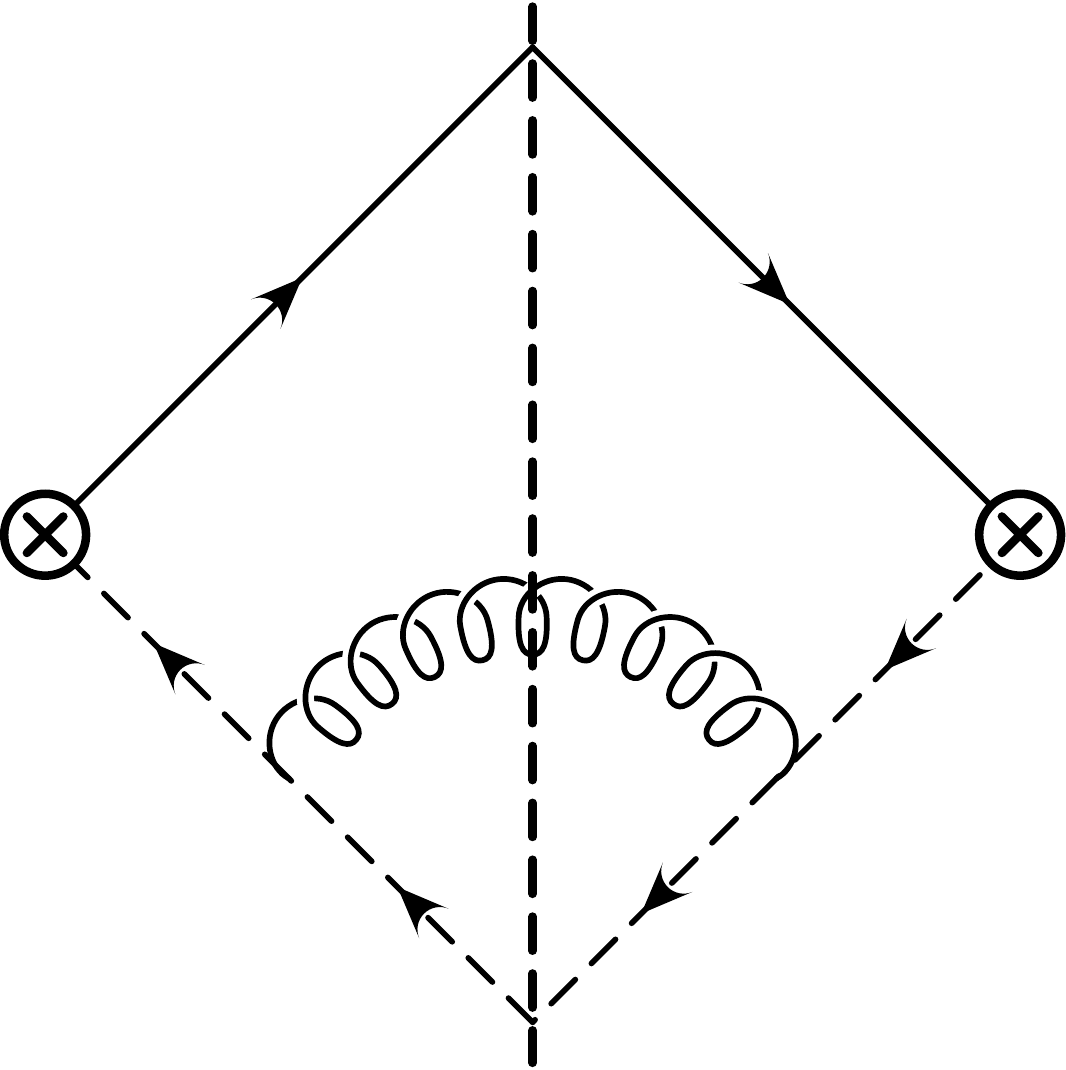}}\hspace{0.05\textwidth}
	\subfloat[$\vac{J^{(0)}}$\label{fig:j0qcd}]{\includegraphics[width=0.15\textwidth]{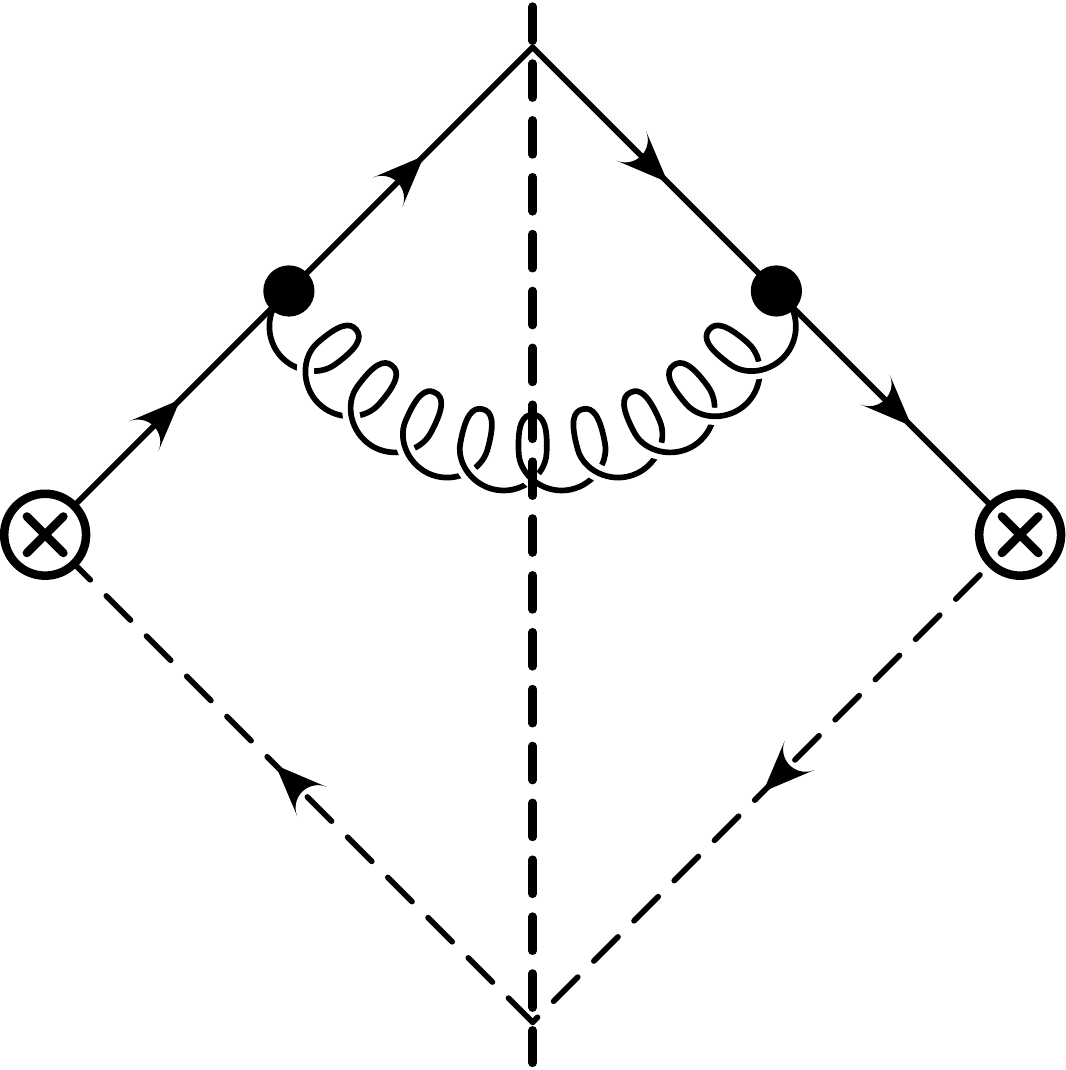}}\\ 
	\vspace{1\baselineskip}
	\subfloat[$\vac{S^{(0)}}$\label{fig:s01}]{\begin{overpic}[width=0.15\textwidth]{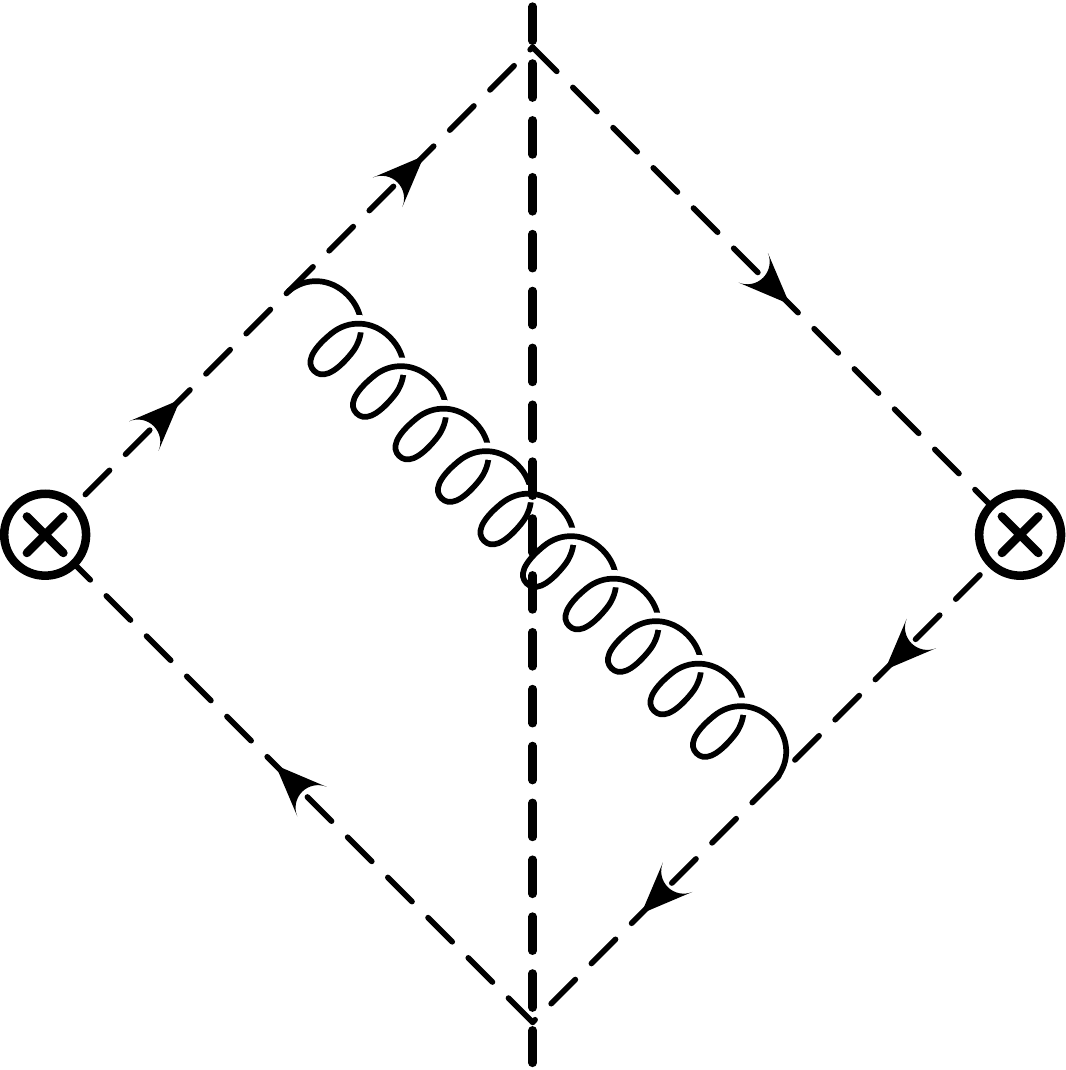}
		\put(15,75){$n$}
		\put(75,75){$n$}
		\put(75,20){$\nb$}
		\put(15,20){$\nb$}
	\end{overpic}
	}\hspace{0.05\textwidth}
	\subfloat[$\vac{S^{(0)}}$\label{fig:s00}]{\includegraphics[width=0.15\textwidth]{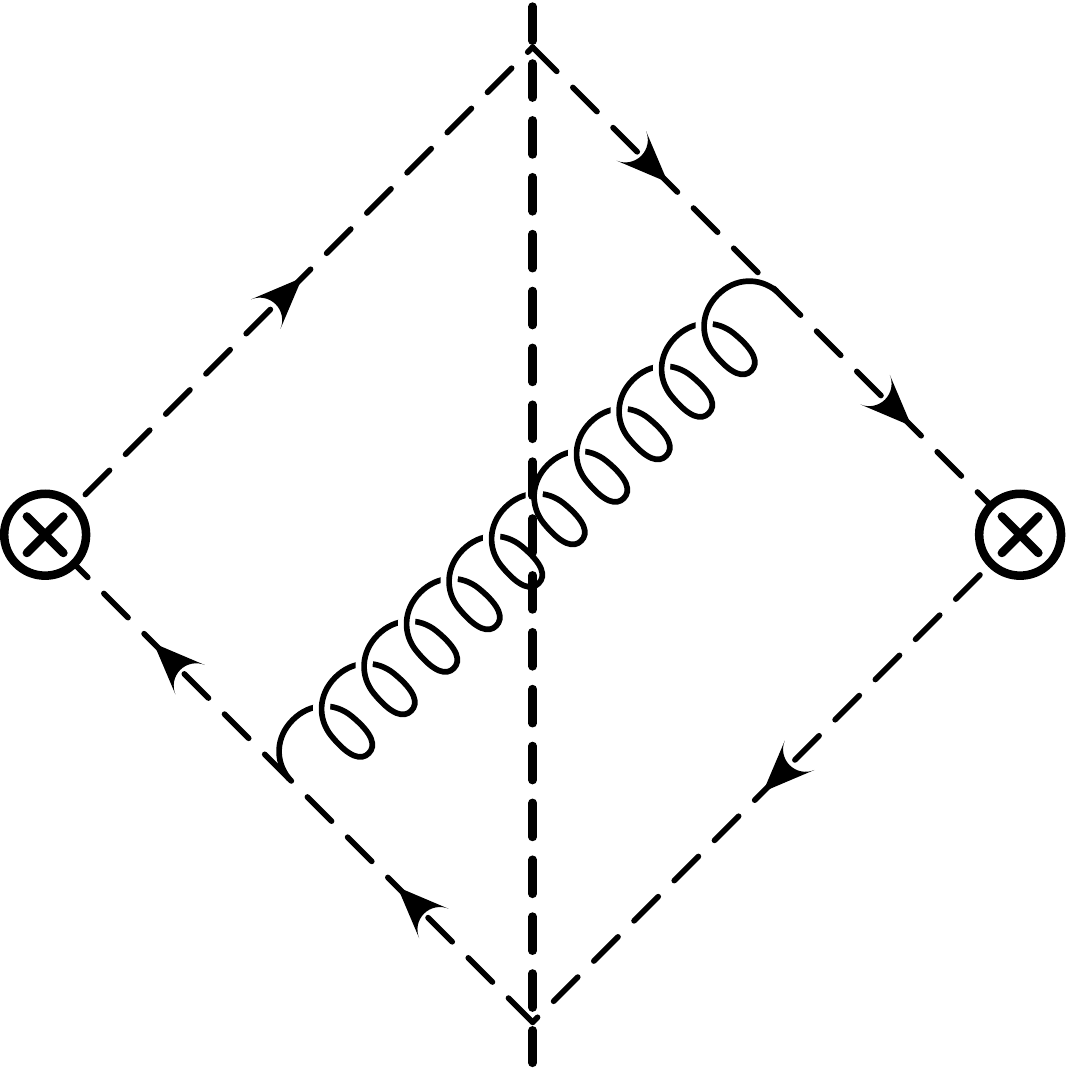}}\hspace{0.05\textwidth}
	\subfloat[$\vac{S^{(0)}}$\label{fig:s03}]{\includegraphics[width=0.15\textwidth]{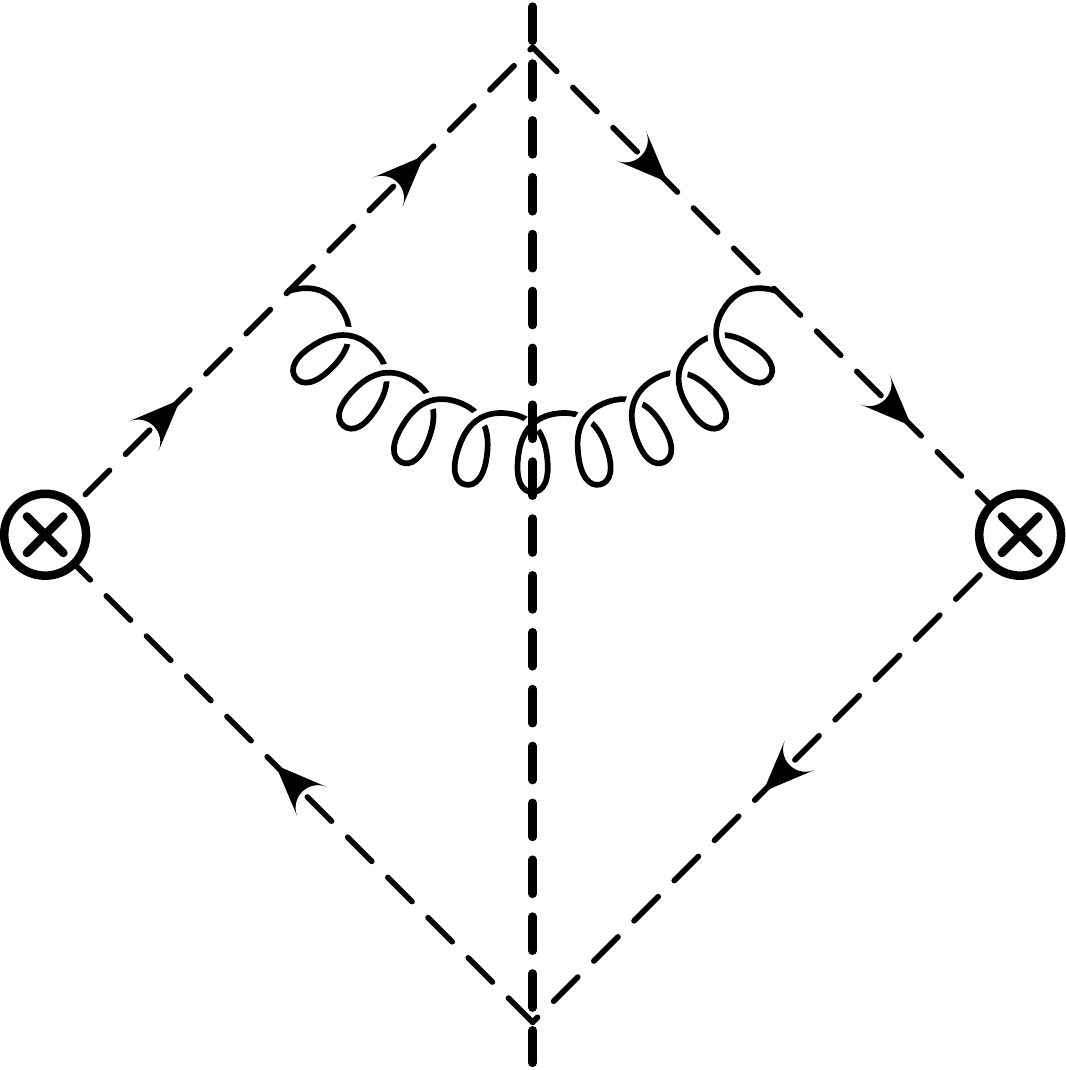}}\hspace{0.05\textwidth}
	\subfloat[$\vac{S^{(0)}}$\label{fig:s04}]{\includegraphics[width=0.15\textwidth]{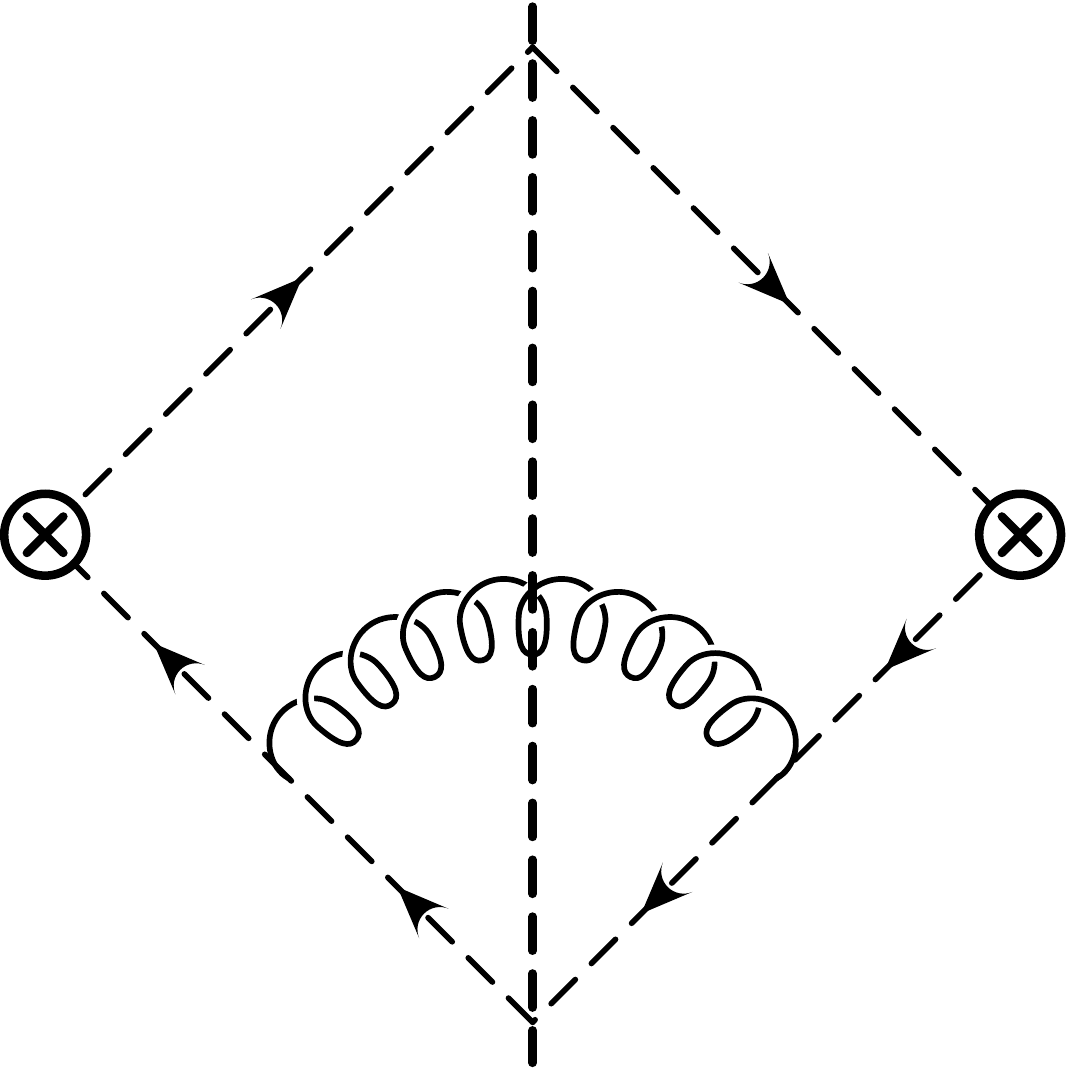}}
	\fcaption{One-loop diagrams of \eqn{JJSlovev}. Solid lines represent fermions, dashed lines represent Wilson lines with the colour flowing in the direction of the arrows, and the dots represent Lagrangian insertions.  The type of soft Wilson lines are labelled in \fig{s01}.  The cut is distinguished by the bold vertical dashed line.  The contributions to $\bar J^{(0)}$ look identical to the contributions of $J^{(0)}$ after a rotation of $180^\circ$.  \label{fig:LO}}
\end{figure*}

By matching the QCD operators in \eqn{rate} onto the SCET operators, the LO thrust rate is written as
\ba\label{SCETrate}
	R(\tau)&=&|C_2^{(0)}|^2 \int d^dx e^{-iQ\cdot x}\\
		&&\times\vac{O_2^{(0)\mu\dagger}(x)\hat\M^{(0)}(\tau)O_{2}^{(0)}\mu(0)} +O(\tau)\non
\ea
where the $O(\tau)$ corrections will be calculated by the subleading in $\la$ operators. The rate is factorized by matching above the operator product onto jet and soft operators
\ba\label{OPElo}
	&&\int d^dx\, e^{-iQ\cdot x}O_2^{(0)\mu\dagger}(x)\hat\M^{(0)}(\tau)O_{2\mu}^{(0)}(0)\nn
	&&=C^{(0)} J^{(0)}(\tau_n)\otimes \bar J^{(0)}(\tau_\nb)\otimes S^{(0)}(\tau_s)
\ea
with matching coefficient $C^{(0)}$.  As usual, the superscripts on the jet and soft operators refer to their suppression in $\la$.  The rate can then be written in the desired factorized form
\ba\label{R000}
	R(\tau)&=&H^{(0)}(\mu)\vac{J^{(0)}(\mu,\tau_n)} \\
		&&\otimes\vac{\bar J^{(0)}(\mu,\tau_\nb)}\otimes\vac{ S^{(0)}(\mu,\tau_s)} +O(\tau)\non
\ea
where the hard function is the product of the matching coefficients, $ H^{(0)}(\mu)=|C_2^{(0)}(\mu)|^2C^{(0)}(\mu)$.

The explicit decoupling of \ncoll, \nbcoll, and soft degrees of freedom in the dijet and measurement operators makes finding the appropriate jet and soft operators straightforward.  Using the Fierz identity to separate the spin and colour indices, we find the operators are
\bwtxt\ba\label{JJSlo}
	J^{(0)}(\mu,\tau)&=&\frac{1}{N_C}\Tr\int \dn\bar\psi_n(0)\wnfun(0,\xsnbinf)\frac{\nbslash}{2}\hat\M_n^{(0)}(\tau)\wnfun(\xnbinf,\xnb)\psi_n(\xnb)\nn
	\bar J^{(0)}(\mu,\tau)&=&\frac{1}{N_C}\Tr\int \dnb\bar\psi_\nb(\xn)\wnbfun(\xn,\xninf)\frac{\nslash}{2}\hat\M_\nb^{(0)}(\tau)\wnbfun(\xsninf,0)\psi_\nb(0)\nn
	S^{(0)}(\mu,\tau)&=&\frac{1}{N_C}\Tr\, \ynbfun(\xsnbinf,0)\ynfun(0,\xsninf)\hat\M_s^{(0)}(\tau)\ynfun(\xsninf,0)\ynbfun(0,\xsnbinf),
\ea
where the trace is over spins and colour.  These operators give the same Feynman rules as those found in \cite{Hornig:2009vb}.  The matching coefficient in \eqn{OPElo} is most easily found by comparing the vacuum expectation value of both sides.   The real emission contributions to the one-loop vacuum expectation value of $J^{(0)}$ and $S^{(0)}$ are pictured in \fig{LO}.  They are calculated by cutting the diagrams along the vertical dashed lines and applying the measurement operator to the fields passing through this cut \cite{Hornig:2009vb}.  The virtual diagrams are scaleless and thus zero in \msbar.  The vacuum expectation value of the jet and soft operators are \cite{Hornig:2009vb}
\ba\label{JJSlovev}
		\vac{J^{(0)}(\mu,\tau)}&=&\delta(\tau) \bigg [ 1+ \alsb \bigg(\frac2{\ep^2}+\frac2\ep\LQ + \frac3{2\ep}+ \LQ^2 + \frac{3}{2} \LQ+ \frac72-\frac{\pi^2}{2} \bigg) \bigg ] +\alsb \left [ \bigg( -\frac2\ep-\frac{3}{2} -2\Ln\bigg )\bigg ( \frac{\theta(\tau)}{\tau}\bigg )\right ]_+ \nn
	\vac{\bar J^{(0)}(\mu,\tau)}&=&\vac{J^{(0)}(\mu,\tau)}\\
	\vac{S^{(0)}(\mu,\tau)}&=&\delta(\tau) \left [1+ \alsb \left (- \frac2{\ep^2}-\frac2\ep\LQ-\LQ^2 +\frac{\pi^2}{6} \right ) \right ]+\alsb \left [ \left(\frac4\ep+4\Ls\right) \left(\frac{\theta(\tau)}{\tau}\right)\right ]_+\nonumber,
\ea\ewtxt
where we have included the zero-bin procedure \cite{Manohar:2006nz}, which accounts for the double counting between the collinear and soft operators.  The matching coefficient $C^{(0)}(\mu)=1$ \cite{Hornig:2009vb} meaning the hard function is
\ba\label{H0}
	H^{(0)}(\mu)&=&1+\alsb\left(\LQ^2-3\LQ-8+\frac{7\pi^2}{6}\right)+O(\al_s^2).\nn
\ea
As expected, the jet and soft operators separate the $\sqrt\tau Q$ and $\tau Q$ scales. The hard function describes the physics above the cut-off $Q$, of the effective theory. 

Matrix elements of the jet and soft operators also do not need any further expansion in $\tau$.  This is required to ensure that operators of different orders do not mix.  We note that this is different than the results in \cite{Cheung:2009sg}, which considered the  exclusive JADE two-jet rate at LO.  This rate has the same $O(\al_s)$ phase space as the thrust rate in QCD.  In \cite{Cheung:2009sg}, the phase space was not consistently expanded in $\la$ and an expansion in $\tau$ was required after the phase space integration.  A subleading zero-bin procedure is necessary to ensure that the LO operators do not contribute to subleading corrections.  In this paper and in \cite{Hornig:2009vb}, the measurement operator is consistently expanded in $\la$ so matrix elements of the jet and soft operators automatically have consistent power counting.

The thrust rate at $O(\al_s \tau^0)$ is calculated by substituting the hard function and the vacuum expectation values of the jet and soft operators into \eqn{R000}.  The rate is found to be
\ba
	R(\tau)&=&1+\alsb\left(-2 \log ^2\tau-3 \log\tau+\frac{\pi ^2}{3}-1\right)+O(\al_s^2,\tau),\nn
\ea
which reproduces the rate found in perturbative QCD at this order \cite{Hornig:2009vb}.  By separately renormalizing the jet and soft operators the large $\log\tau$'s were summed in \cite{Hornig:2009vb}.    In the next section we follow the same procedure to find the $O(\al_s\tau)$ correction to the rate and write it in a factorized form analogous to \eqn{R000}.


\section{Next-to-Leading Order Calculation\label{sec:nlo}}

The results are extended to include the $O(\tau)$ corrections by systematically matching the QCD current and measurement operator in \eqn{rate} onto subleading SCET dijet and measurement operators.  The $O(\tau)$ thrust rate in SCET is written as
\ba\label{SCETratenlo}
	R(\tau)&=&\sum_{i+j+k\leq2}C_2^{(i)*} C_2^{(j)}\int d^4xe^{-iQ\cdot x}\\
		&&\times\vac{O_2^{(i)\dagger\mu}(x)\hat\M^{(k)}(\tau)O_{2\mu}^{(j)}(0)}+O(\tau^2)\nonumber.
\ea
From the LO calculation $\la\sim\sqrt{\tau}$, so we need the $O(\la^2)$ SCET operators as illustrated by the constraints on the sum.  One can explicitly check  that the $O(\la)$ corrections, which would give $O(\sqrt{\tau})$ corrections, vanish.

 As in the previous section, we want to write \eqn{SCETratenlo} in a factorized form by matching onto subleading jet and soft operators.  These operators will be generalizations of the LO operators and their matrix elements must have a consistent power counting in $\tau$ as in the LO case.  We will only do the tree-level matching, which is all that is necessary to calculate the $O(\al_s\tau)$ rate.  The subleading dijet operators are written in \app{operators} and explicitly decouple the sectors.  We must also find the subleading measurement operators, $\hat\M^{(1,2)}(\tau)$ in \eqn{Mmatch}.  We will show in the next section that the action of the subleading measurement operators also decouples the sectors, analogously to \eqn{Mfact}.  The explicit decoupling of the sectors in the dijet and measurement operators makes writing the rate in the desired factorized form in \sec{fact} straightforward.


\subsection{Measurement Operator}

The action of the subleading measurement operators are found by first expanding the definition of the thrust axis and then finding this expansion's effect on the measurement of thrust.  The thrust axis is defined as the unit vector that maximizes the sum below \eqn{thrust}.  The sum is maximized when the thrust axis is in the direction of the hemisphere with the largest three-momentum,  $\vec p_{+t}$ \cite{Bauer:2008dt}.  Therefore, the thrust axis is written in SCET as the expansion 
\ba\label{tqcd}
	\vec t=\frac{\vec p_{+t}}{|\vec p_{+t}|}=\vec t^{\,(0)}+\vec t^{\,(2)}+\vec t^{\,(4)}+O(\lambda^6),
\ea  
where the superscripts refer to the suppression in $\la$. 

In order to find $\vec p_{+t}$, we first note that the \ncoll\ and \nbcoll\ particles are always in opposite hemispheres.  SCET momentum power counting enforces this at LO and the zero-bin procedure will enforce this at all orders in $\la$.  Therefore, the total momentum of the hemisphere will be
\ba\label{P+t}
	\vec p_{+t}=\vec p_\nb+\vec k_{+t},
\ea
where $p_\nb^\mu$ is the total \nbcoll\ momentum and $k_{+t}^\mu$ is the total soft momentum in the $+\vec t$ hemisphere as defined in \eqn{khemi}.  The expansion of \eqn{khemi} and \eqn{tqcd} allows us to iteratively solve for both $\vec t^{\,(i)}$ and $k^{(i)\mu}_{\pm t}$.  The subleading corrections to the thrust axes are
\ba\label{tnlo}
	\vec t^{\,(2)}&=&  \frac{2\kAp}{p_\nb^+}\\
	\vec t^{\,(4)}&=&\frac{2 (\kAp)^2}{(p_\nb^+)^2}\vec n +\frac{(2\vec n\cdot \vec k_{+t}^{(0)}+k_\nb^-)}{(p_\nb^+)^2}\kAp+\frac{2\kApnlo}{p_\nb^+}\nonumber
\ea
where $k_\nb^-$ is the total \nbcoll\ momentum in the $n^\mu$ direction.  The subleading correction to the total soft momentum in each hemisphere
\ba\label{kAnlo}
	 k^{(2)\mu}_{+ t}=- k^{(2)\mu}_{- t}= \frac{1}{p_\nb^+}\sum_{i}k_i^\mu\left(2\kAp\cdot \vec k_{i\perp}\right)\de(-\vec n\cdot \vec k_i)\nn
\ea
is found by inserting the $\vec t^{\,(2)}$ into \eqn{khemi}.  The first equality above is because the sum of the soft momentum in the two hemispheres must be $O(\la^2)$.  

We note we could have instead used $-\vec p_{+t}=\vec p_n+\vec k_{-t}$ in \eqn{tqcd}, where $\vec p_n$ is the total \ncoll\ momentum and  $\vec k_{-t}$ is the total soft momentum in the $-\vec t$ hemisphere.  However, the definition in \eqn{P+t} is simpler due to our choice of $\vec n$ in \sec{lo}  such that $\vec p_{\nb\perp}=\vec 0$.  The apparent asymmetry in the labelling of $n^\mu$ and $\nb^\mu$ in the resulting phase space will be accounted for by the $O_2^{(1\de)}$ and $O_2^{(2\de_\perp)}$ operators in \app{operators}.  The choice of $\vec p_{\nb\perp}\equiv\vec 0$ means the Feynman rules of these operators involve $\partial/\partial \vec p_{n\perp}$'s only.

The subleading measurement operators are found by substituting the corrections to the thrust axis into \eqn{thrust}.  We first consider the contribution from an \ncoll\ particle with momentum $p_i$.  As discussed in the second paragraph of this section, $E_i+\vec p_i\cdot \vec t$ is always the minimum for each \ncoll\ particle.  Therefore, the \ncoll\ sector contributes
\ba\label{nthrust}
	\frac1Q\sum_i(E_i+\vec p_i\cdot\vec t^{\,(0)})+\frac{1}{Q}&&\left(\vec t^{\,(2)}+\vec t^{\,(4)}\right)\cdot \sum_i \vec p_{ i}+O(\lambda^5)\nn
\ea
to the thrust, where the sum is only over \ncoll\ particles.  The first term reproduces the action of the LO measurement operator in \eqn{Mlocs}.  The $t^{(2)}$ term gives the NLO correction and the first term of $t^{(4)}$ in \eqn{tnlo} gives the \nnlo\ correction.  This power counting is due to $p_{\perp i}\sim O(\la)Q$ for \ncoll\ particles. 

The contribution to thrust from the \nbcoll\ particles is found in a similar way.  Here, the $E_i-\vec p_i\cdot\vec t$ is always the minimum so the \nbcoll\ sector contributes
\ba\label{nbthrust}
	\frac1Q\sum_i(E_i-\vec p_i\cdot\vec t^{\,(0)})-\frac{1}{Q}&&\vec t^{\,(4)}\cdot \sum_i \vec p_{ i}+O(\lambda^5),
\ea
where the sum is only over \nbcoll\ particles.  The first term is the LO contribution and the second term is the \nnlo\ contribution.  There is no $\vec t^{\,(2)}$ term because we have set $\vec p_{\nb\perp}=\vec0$ by our choice of $\vec n$.  

Unlike the collinear particles, soft particles can be in either hemisphere.  The minimum of $E_i\pm\vec t\cdot \vec k_i$ is determined by which hemisphere the soft particle is in.  Therefore, the soft sector contributes
\ba\label{sthrust}
	&&\frac{1}{Q}\left(n\cdot \kA+\nb\cdot\kB\right)\\
	&&+\frac1Q\left(n\cdot \kAnlo+\nb\cdot\kBnlo+\vec t^{\,(2)}\cdot\left(\vec k_{+ t}^{(0)}-\vec k_{- t}^{(0)}\right) \right)+O(\lambda^6).\non
\ea
to the total thrust.  The first line is the LO contribution in \eqn{Mlocs} and the remaining terms are all \nnlo.

The action of the subleading measurement operators is found by Taylor expanding the contribution to thrust in $\la$.  We incorporate the NLO and \nnlo\ corrections from the collinear sectors in \eqn{nthrust} and \eqn{nbthrust} by writing the action of the subleading measurement operators as
\bwtxt\ba\label{MNnlo}
	\M^{(1n)}(\tau, \{p_n,p_\nb,k_s\})&=&\left(\frac{p^\al_{n\perp}}{Q}\frac{\partial}{\partial \tau_n}\M_n^{(0)}(\tau_n,\{p_n\})\right)\otimes\left(\frac{Q}{p_\nb^+}\M_\nb^{(0)}(\tau_\nb,\{p_\nb\})\right)\otimes\left(\frac{-2k^{(0)\al}_{+ t\perp}}{Q}\M_s^{(0)}(\tau_s,\{k_s\})\right)\nn
	\M^{(2n_a)}(\tau, \{p_n,p_\nb,k_s\})&=&\left(\frac{p^\al_{n\perp}p^\beta_{n\perp}}{Q^2}\frac{\partial^2}{\partial \tau_n^2}\M_n^{(0)}(\tau_n,\{p_n\})\right)\otimes\left(\frac{Q^2}{p_\nb^{+2}}\M_\nb^{(0)}(\tau_\nb,\{p_\nb\})\right)\otimes\left(\frac{2k^{(0)\al}_{+ t\perp}k^{(0)\beta}_{+ t\perp}}{Q^2}\M_s^{(0)}(\tau_s,\{k_s\})\right)\non
\ea
\ba
	\M^{(2n_b)}(\tau, \{p_n,p_\nb,k_s\})&=&\left(\frac{p^-_{n}}{Q}\frac{\partial}{\partial \tau_n}\M_n^{(0)}(\tau_n,\{p_n\})\right)\otimes\left(\frac{Q^2}{(p_\nb^+)^2}\M_\nb^{(0)}(\tau_\nb,\{p_\nb\})\right)\otimes\left(\frac{-(\kAp)^2}{Q^2}\M_s^{(0)}(\tau_s,\{k_s\})\right)\\
	\M^{(2\nb_b)}(\tau, \{p_n,p_\nb,k_s\})&=&\M_n^{(0)}(\tau_n,\{p_n\})\otimes\left(\frac{p_\nb^+}{Q}\frac{\partial}{\partial \tau_\nb}\M_\nb^{(0)}(\tau_\nb,\{p_\nb\})\right)\otimes\left(\frac{(\kAp)^2}{Q^2}\M_s^{(0)}(\tau_s,\{k_s\})\right).\nonumber
\ea\ewtxt
$\M^{(2n_a)}$ comes from the second term in the expansion of the NLO correction in \eqn{nthrust}.   The \nnlo\ corrections from the soft sector in \eqn{sthrust} are incorporated by the subleading measurement operators
\ba\label{MSnlo}
	&&\M^{(2s_{i})}(\tau, \{p_n,p_\nb,k_s\})=\\
	&&\M_n^{(0)}(\tau_n,\{p_n\})\otimes\left(\frac{Q}{p_\nb^+}\M_\nb^{(0)}(\tau_\nb,\{p_\nb\})\right)\otimes\M_s^{(2s_i)}(\tau_s,\{k_s\})\non
\ea
where
\ba\label{MSi}
	\M_s^{(2s_{1})}(\tau, \{k_s\})&=&\frac{2\kAp\cdot\kAp}{Q}\frac{\partial}{\partial \tau_s}\M_s^{(0)}(\tau_s,\{k_s\})\nn
	\M_s^{(2s_{2})}(\tau, \{k_s\})&=&\frac{-2\kAp\cdot\kBp}{Q}\frac{\partial}{\partial \tau_s}\M_s^{(0)}(\tau_s,\{k_s\})\nn
	\M_s^{(2s_{- t})}(\tau, \{k_s\})&=&\frac{p_\nb^+n\cdot\kAnlo}{Q^2}\frac{\partial}{\partial\tau_s}\M_s^{(0)}(\tau_s,\{k_s\})\\
	\M_s^{(2s_{+ t})}(\tau, \{k_s\})&=&\frac{p_\nb^+\nb\cdot\kBnlo}{Q^2}\frac{\partial}{\partial\tau_s}\M_s^{(0)}(\tau_s,\{k_s\})\nonumber.
\ea
The $p_\nb^+$ in the action of the soft measurement operators $\M^{(2s_{\pm t})}$ are introduced to cancel the $p_\nb^+$ in \eqn{kAnlo}.  

The actions of the measurement operators \eqn{MNnlo} and \eqn{MSnlo} define the NLO and \nnlo\ measurement operators $\hat M^{(1,2)}(\tau)$.  As the brackets suggest, the sectors explicitly decouple in the action of the subleading measurement operators.  This is due to the corrections to thrust depending only on the total momentum of each collinear sector and not any individual particle.  While it is possible to formally write the subleading measurement operators using the energy-flow operator \cite{Bauer:2008dt, Hornig:2009vb} and not just their actions in momentum space, we see no reason to do so: only their actions in momentum space are necessary in calculations.

We note that the measurement operators in this section are found using the formalism of \cite{Freedman:2011kj} and would be different if we used the SCET formalism of \cite{\labelscet}.  It was suggested in \cite{Bauer:2008dt} the subleading measurement operators could be found using the subleading terms in the SCET Lagrangian of \cite{\labelscet}.  While we do not explicitly check this, we note that the breaking of the explicit decoupling of soft and collinear fields in the subleading Lagrangian would complicate factorization.


\subsection{Factorization \label{sec:fact}}

The explicit decoupling of the sectors in the subleading dijet and measurement operators makes it straightforward to factorize the subleading corrections to the rate.  In order to factorize the rate, each operator product in \eqn{SCETratenlo} is matched onto the appropriate jet and soft operators
\ba\label{ope}
	&&\int d^dx\, O_2^{(i)\dagger}(x)\hat\M^{(k)}(\tau)O_2^{(j)}(0)\\
		&&=\sum_{\al(l,m,n)}C^{(i,j,k)}_{\al} J^{(l)}(\tau_n)\otimes\bar J^{(m)}(\tau_\nb)\otimes S^{(n)}(\tau_s)\nn
		&&+\tau\, C^{(i,j,k)}_{0} J^{(0)}(\tau_n)\otimes\bar J^{(0)}(\tau_\nb)\otimes S^{(0)}(\tau_s)\nonumber,
\ea
with matching coefficients $C_{\al,0}^{(i,j,k)}$.  The last line is made \nnlo\ by the explicit $\tau$ in front.  The integer $\al\geq1$ labels the combination of jet and soft operators and we require  $i+j+k=2=l+m+n$.  The $O(\tau)$ rate can then be written in the factorized form
\bwtxt\ba\label{Rijklmn}
	R^{(2)}(\tau)=&&\sum_{i,j,k,\al}H_{\al}^{(i,j,k)}(\mu)\vac{J^{(l)}(\mu,\tau_n)}\otimes\vac{\bar J^{(m)}(\mu,\tau_\nb)}\otimes \vac{S^{(n)}(\mu,\tau_s)}\nn
	&&+\tau\,\sum_{i,j,k}H_{0}^{(i,j,k)}(\mu)\vac{J^{(0)}(\mu,\tau_n)}\otimes\vac{\bar J^{(0)}(\mu,\tau_\nb)}\otimes \vac{S^{(0)}(\mu,\tau_s)}
\ea\ewtxt
where the hard functions are defined as
\ba
	H_{\al,0}^{(i,j,k)}(\mu)&=&C_2^{(i)*}(\mu)C_2^{(j)}(\mu)C_{\al,0}^{(i,j,k)}(\mu) .
\ea
This generalizes the LO factorization \eqn{R000} to incorporate the \nnlo\ corrections and is the main result of our paper.  While it is possible to calculate the $O(\al_s\tau)$ rate directly from \eqn{SCETratenlo}, the factorized form will allow the jet and soft operators to be renormalized separately.  

Below we will use a few examples to demonstrate how the jet and soft operators in \eqn{ope} are found.  The full list of operators and their matching coefficients are found in \app{jjs}.  For the sake of brevity, we only write those operators that contribute to the $O(\al_s\tau)$ rate.  We omit the phase space integrals when calculating matrix elements of the operators to avoid overbearing the reader.

\begin{figure}[!t]
	\centering
	\subfloat[\label{fig:j2abn1}]{\includegraphics[width=0.15\textwidth]{j0L.pdf}}
	\hspace{0.005\textwidth}
	\subfloat[\label{fig:j2abn2}]{\includegraphics[width=0.15\textwidth]{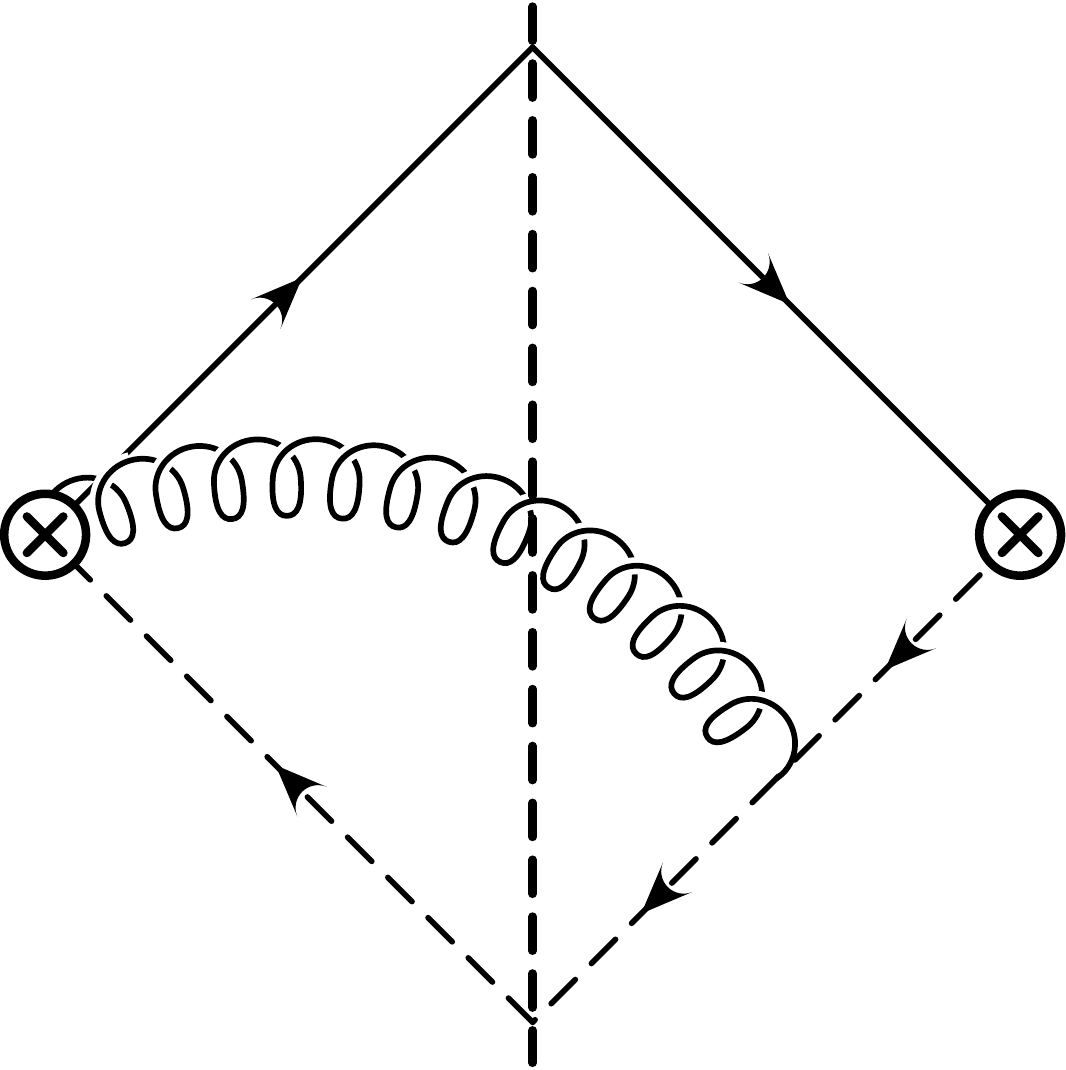}}
	\hspace{0.005\textwidth}
	\subfloat[\label{fig:j2abn0}]{\includegraphics[width=0.15\textwidth]{j0nn.pdf}}
	\fcaption{Matrix element $\vac{J^{(2ab_n)}}$ that give non-zero $\tau$ values at $O(\al_s)$.\label{fig:j2abn}}
\end{figure}

\parsec{i=1a_n,\, j=1b_n,\, k=0} The operators are found in \eqn{O21n}.  The left-hand side of \eqn{ope} is 
\ba\label{1an1bn}
	&&-\int d^d x e^{-iQ\cdot x}\frac1Q\left[\bar\psi_\nb(\xn)\wnbfun(\xn,\xninf)\right]\nn
		&&\times\left[\ynbfun(\xsnbinf,0)\ynfun(0,\xsninf)\right]\nn
		&&\times\left[\wnfun(\xnbinf,\xnb)i\overleftarrow D^{\beta_1}_\perp(\xnb)\Gamma_1^{\beta_1\mu}\psi_n(\xnb)\right]\nn
		&&\times\hat\M_n^{(0)}(\tau_n)\otimes\hat\M_\nb^{(0)}(\tau_\nb)\otimes \hat\M_s^{(0)}(\tau_s)\\
		&&\times\frac1Q\left[\bar\psi_n(0)i\overleftarrow{D}^{\beta_2}_\perp(0)\Gamma_2^{\beta_2\mu}\wnfun(0,\xsnbinf)\right]\nn
		&&\times\left[\ynfun(\xsninf, 0)\ynbfun(0,\xsnbinf)\right]\left[\wnbfun(\xsninf,0)\psi_\nb(0)\right]\nonumber
\ea
where $\Ga_1^{\al\mu}=\frac{\nslash}2\gamma^\al_\perp\gamma^\mu P_n$ and $\Ga_2^{\al\mu}=\frac{\nbslash}{2}\gamma^\al_\perp\gamma^\mu P_\nb$ and $\ga^\mu_\perp$ is defined in \app{jjs}.  The $1/Q$'s come from the definition of the operator expansion \eqn{qcdmatch} and the negative sign comes from taking the hermitian conjugate.  The square brackets denote the separately gauge invariant pieces of each sector.  As for the LO factorization, we use the Fierz identity to separate the spin and colour indices and match onto the appropriate jet and soft operators
\ba\label{ope1an1bn0}
	&&C_{1}^{(1a_n,1b_n,0)}J^{(2ab_n)}(\tau_n)\otimes\bar J^{(0)}(\tau_\nb)\otimes S^{(0)}(\tau_s)\nn
	&&+\tau\, C_{0}^{(1a_n,1b_n,0)} J^{(0)}(\tau_n)\otimes\bar J^{(0)}(\tau_\nb)\otimes S^{(0)}(\tau_s).
\ea
The LO operators $J^{(0)}, \bar J^{(0)}$ and $S^{(0)}$ are defined in \eqn{JJSlo} and the subleading jet operator is
\bwtxt\ba\label{J1an1bn}
	J^{(2ab_n)}(\mu,\tau)&=&\frac1{Q^2N_C}\Tr\int\dn \bar\psi_n(0)iD_\perp^\al(0)\wnfun(0,\xsninf)\frac{\nbslash}{2}\hat\M_n^{(0)}(\tau)\wnfun(\xnbinf,\xnb)iD_{\perp\al}(\xnb)\psi_n(\xnb).\nn
\ea\ewtxt
The operator is suppressed by $\lambda^2$ due to the derivative insertions. The vacuum expectation value of this operator is shown in \fig{j2abn}. 

In order to find the matching coefficients, we calculate the vacuum expectation value of \eqn{1an1bn} and \eqn{ope1an1bn0}.  The Feynman rules for the dijet operators were written in \cite{Freedman:2011kj}, and we find the vacuum expectation value of \eqn{1an1bn} is 
\ba
	-\frac12\alsb\tau+O(\al_s^2).
\ea
The vacuum expectation values of $J^{(0)}$, $\bar J^{(0)}$, and $S^{(0)}$ were found in \eqn{JJSlo} and the diagrams in \fig{j2abn} lead to
\ba
	\vac{J^{(2ab_n)}(\mu,\tau)}=\frac12\alsb+O(\al_s^2).
\ea
Here and in the calculations below we have included the zero-bin procedure.  As expected, the matrix element of the $J^{(2ab_n)}$ operator is suppressed by $\tau\sim\la^2$ compared to the LO jet operator.   The matching coefficients are found to be $C_{1}^{(1a_n,1b_n,0)}=-1+O(\al_s)$ and $C_{0}^{(1a_n,1b_n,0)}=0+O(\al_s^2)$ meaning the hard functions are
\ba
	H_{1}^{(1a_n,1b_n,0)}(\mu)&=&1+O(\al_s)\nn
	H_{0}^{(1a_n,1b_n,0)}(\mu)&=&0+O(\al_s^2).
\ea
Therefore, the contribution of \eqn{1an1bn} to the rate can be written in the factorized form \eqn{Rijklmn}, with the appropriate subleading jet operator.

\parsec{i=2\de_n,\,j=k=0}  The operator $O_2^{(2\de_n)}$ is shown in \eqn{O22n} and accounts for the matching of QCD momentum conservation onto SCET momentum conservation at $O(\la^2)$.  The contribution where $j=2\de_n$ in \eqn{ope} means taking $O_2^{(2\de_n)}(0)$, which vanishes due to the explicit $x$ dependence in the dijet operator.  As in the previous example, we use the Fierz identity to factorize the contribution from this operator and match onto 
\ba\label{OPE2den0}
	&&\int d^dx\, O_2^{(2\de_n)\dagger}(x)\hat\M^{(0)}(\tau)O_2^{(0)}(0)\\
	&=&C^{(2\de_n,0,0)}_{1}J^{(2\de_n)}(\tau_n)\otimes\bar J^{(0\de_n)}(\tau_\nb)\otimes S^{(0)}(\tau_s)\nn
		&&+\tau\,C^{(2\de_n,0,0)}_{0} J^{(0)}(\tau_n)\otimes\bar J^{(0)}(\tau_\nb)\otimes S^{(0)}(\tau_s)\nonumber,
\ea
where the new jet operators are
\bwtxt\ba\label{J2den}
	J^{(2\de_n)}(\tau)&=&\frac{1}{Q N_C}\Tr\int\dn\bar\psi_n(0)\wnfun(0,\xsninf)\frac{\nbslash}{2}\hat\M^{(0)}_n(\tau) \wnfun(\xnbinf,\xnb)\left(in\cdot \overleftarrow{D}+in\cdot D\right)(\xnb)\psi_n(\xnb)\nn
	\bar J^{(0\de_n)}(\tau)&=&\frac{iQ}{N_C}\Tr\int\dnb\, x^-\bar\psi_\nb(\xn)\wnbfun(\xn,\xninf)\frac{\nslash}{2}\hat\M_\nb^{(0)}(\tau)\wnbfun(\xsnbinf,0)\psi_\nb(0).
\ea\ewtxt
The derivatives in $J^{(2\de_n)}$ pull down the $O(\la^2)$ component of the \ncoll\ momentum leading to a $\la^2$ suppression compared to the LO jet operator.  The $x^-$ dependence of $O_2^{(2\de_n)}$ is put into $\bar J^{(0\de_n)}$ because only the \nbcoll\ fields depend on this coordinate.  We have introduced the $Q$'s so the operators will have the correct mass dimensions.  The diagrams of the vacuum expectation value of $J^{(2\de_n)}$ and $\bar J^{(0\de_n)}$ are the same as the diagrams of the vacuum expectation value of $J^{(0)}$ and $\bar J^{(0)}$ in \fig{LO}.   

As before, the matching coefficient is most easily determined by comparing the vacuum expectation value of both sides of \eqn{OPE2den0}.  The explicit $x^-$ in both $O_2^{(2\de_n)}$ and $\bar J^{(0\de_n)}$, which is a derivative $\partial/\partial p_\nb^+$ in momentum space, acts on the $d$-dependent prefactor in $\M_\nb^{(0)}$ of \eqn{Mlocs}.  Therefore, the vacuum expectation value of the left-hand side of \eqn{OPE2den0} is
\ba\label{2den0}
	\alsb\tau\left(-\frac4\ep-4\Ln-3\right)+O(\al_s^2).
\ea
and the vacuum expectation value of the jet operators are
\ba\label{J2den0den}
	\vac{J^{(2\de_n)}(\mu,\tau)}&=&\alsb\left(-\frac2\ep-2\Ln-\frac32\right)+O(\al_s^2)\nn
	\vac{\bar J^{(0\de_n)}(\mu,\tau)}&=&2\de(\tau)+O(\al_s).
\ea
Comparing both sides of \eqn{OPE2den0} we find the matching coefficients are $C_{1}^{(2\de_n,0,0)}(\mu)=1+O(\al_s)$ and $C_{0}^{(2\de_n,0,0)}(\mu)=\alsb+O(\al_s^2)$.  We see in this example why the last line of \eqn{ope} is necessary: the $d-2$ from the derivative acting on $\M_n^{(0)}$ gives an extra term in \eqn{2den0} when it is analytically continued to $d=4$ compared to $\vac{\bar J^{(0\de_n)}}$, which has no poles at $d=4$.  The contribution of this operator product is thus factorized in the form of \eqn{Rijklmn} with hard functions
\ba
	H_{1}^{(2\de_n,0,0)}(\mu)&=&-1+O(\al_s)\nn
	H_{0}^{(2\de_n,0,0)}(\mu)&=&-\alsb+O(\al_s^2).
\ea
Again, the matrix elements of the jet operators have the appropriate power counting and the logarithm in \eqn{J2den0den} is minimized at the jet scale, as expected.

\parsec{i=1\de_s,\,j=0,\,k=1n} As a final example, we show an insertion of the subleading measurement operator.  The contribution of the NLO measurement function, $k=1n$, vanishes at $O(\al_s)$ unless $i=1\de_s$.   This contribution accounts for the phase space we neglected at LO due to not conserving soft momentum.  We factorize this operator product by matching
\ba
	&&\int d^dx\, O_2^{(1\de_s)\dagger}(x)\hat\M^{(1n)}(\tau)O_2^{(0)}(0)\nn
	&=&C^{(1\de_s,0,1n)}_{1}J^{(-2\de_s M_n)}(\tau_n)\otimes\bar J^{(0_{M1})}(\tau_\nb)\otimes S^{(4\de_s M)}(\tau_s)\nn
	&&+\tau\,C_{0}^{(1\de_s,0,1n)} J^{(0)}(\tau_n)\otimes\bar J^{(0)}(\tau_\nb)\otimes S^{(0)}(\tau_s).
\ea
The decoupling of the sectors in $\hat\M^{(1n)}$ means the NLO measurement operator can be treated identically as the LO measurement operator in \eqn{OPElo} and gets pulled through when we use the Fierz identity.  Therefore, the appropriate jet and soft operators are
\bwtxt\ba
	J^{(-2\de_sM_n)}(\mu,\tau)&=&\frac{1}{QN_C}\Tr\int\dn\,x_{\perp\al}\bar\psi_n(0)\wnfun(0,\xsninf)\frac{\nbslash}{2}\overleftarrow{\partial}^\al_\perp\frac{\partial}{\partial\tau}\hat\M^{(0)}_n(\tau)\wnfun(\xnbinf,\xnb)\psi_n(\xnb)\nn
	\bar J^{(0_{M1})}(\mu,\tau)&=&\frac{iQ}{ N_C}\Tr\int_0^\infty dt\int\dnb\bar\psi_\nb(\xn)\wnbfun(\xn,\xninf)\frac{\nslash}{2}\hat\M_\nb^{(0)}(\tau)\wnbfun(\xsnbinf,nt)\psi_\nb(nt)\\
	S^{(4\de_sM_n)}(\mu,\tau)&=&\frac{1}{QN_C}\Tr\,\ynfun(\xsninf, 0)  \left(iD_{\perp\al}+i\overleftarrow{D}_{\perp\al}\right)(0)\ynbfun(0, \xsnbinf)\widehat\M_s^{(2n)\al}(\tau)\ynbfun(\xsnbinf,0)\ynfun(0,\xsninf)\nonumber
\ea\ewtxt
where 
\ba
	\M_s^{(2n)\al}(\tau,\{k\})&=&\frac{2k^{(0)\al}_{+t\perp}}{Q}\M_s^{(0)}(\tau,\{k\}).
\ea
The explicit $x_\perp$ dependence in $O_2^{(1\de_s)}$ is put into $J^{(-2\de_sM_n)}$ because we chose the $\vec n$ axis such that $\vec p_{\nb\perp}\equiv \vec0$.  The $p_{n\perp}^\mu$ in $\M_n^{(1n)}$ acts as a total derivative at the cut so becomes a derivative at infinity\footnote{We use a covariant gauge so the gauge field vanishes at infinity.}.  We have also used the identity \cite{Bauer:2001ct,Beneke:2002ph} $(in\cdot\partial)^{-1}\phi(x)=\int_0^\infty dt\,\phi(x+nt)$ to write the $1/p_\nb^+$ in \eqn{MNnlo} as a displacement in the $\bar J^{(0_{M1})}$ operator.  The power counting suggests  $J^{(-2{\de_sM_n})}\sim \lambda^{-2}$ will be enhanced compared to the LO jet operator.  However, it is always convoluted with $S^{(4\de_sM_n)}\sim\lambda^4$, meaning the contribution to the rate will be $O(\la^2)$ as expected.  The diagrams for the vacuum expectation value of the jet and soft operators have the same picture as \fig{LO} and give
\ba\label{JJS1M1de}
	&&\vac{J^{(-2{\de_sM_n})}(\mu,\tau)}=2\de'(\tau)+O(\al_s)\nn
	&&\vac{\bar J^{(0_{M1})}(\mu,\tau)}=\de(\tau)+O(\al_s) \\
	&&\vac{S^{(4{\de_sM_n})}(\mu,\tau)}=-4\alsb\tau +O(\al_s^2),\nonumber
\ea
which have the expected power counting in $\tau$.  The matching coefficients are found to be $C_{1}^{(1{\de_s},0,1{M_n})}(\mu)=1+O(\al_s)$ and $C_{0}^{(1\de_s,0,1M_n)}(\mu)=0+O(\al_s^2)$ and the contribution from this operator can be written in the factorized form of \eqn{Rijklmn} with hard functions
\ba
	H_{1}^{(1\de_s,0,1M_n)}(\mu)&=&1+O(\al_s)\nn
	H_{0}^{(1\de_s,0,1M_n)}(\mu)&=&0+O(\al_s^2).
\ea
This example shows how the decoupling of the measurement operators makes it straightforward to factorize their contribution. 

The factorization of the rest of the subleading dijet and measurement operators follows in the same way as the above examples.  The explicit decoupling of the sectors makes finding the jet and soft operators a matter of using the Fierz identity to separate spinor and colour indices.  The required jet and soft operators are written in \app{jjs}.  These operators are generalizations of the LO operators found in \sec{lo}.  The matching coefficients are found by comparing the vacuum expectation values of \eqn{ope}.  These matrix elements are pictured in \fig{JJSnlo} of \app{jjs} and their values are shown in \tab{vev} of the Appendix.  The appropriate matching coefficients are shown in \tab{match}.  Combining the results of this Table, we find the $O(\al_s\tau)$ rate is
\ba\label{Rnlo}
	R(\tau)=&&1+\alsb\left(-2\log^2\tau-3\log\tau+\frac{\pi^2}{3}-1+\tau(2\log\tau-4)\right)\nn
	&&+O(\al_s^2,\tau^2).
\ea
We can compare these results with those found using perturbative QCD in \cite{Kramer:1986mc}.  Although the results in \cite{Kramer:1986mc} are for the exclusive two-jet rate using the JADE algorithm, the $O(\al_s)$ QCD phase space is the same as the phase space for thrust, as mentioned above.  Therefore, the full $O(\al_s)$ result calculated in \cite{Kramer:1986mc} can be compared to \eqn{Rnlo}.  Summing the $\tau\log\tau$'s requires renormalizing the jet and soft operators, which we do not do here.

\section{Conclusion \label{sec:concl}}

We have shown how to systematically calculate the $O(\tau)$ corrections to the thrust rate in the perturbative regime using SCET.  The rate was factorized and written as the convolution of the vacuum expectation value of jet and soft operators.  Each operator has consistent power counting and depends on a different scale associated with the rate. The appropriate jet and soft operators are found in this work, as well as the matching coefficients.  The $O(\al_s\tau)$ rate was calculated and reproduced the rate found using perturbative QCD.

The rate was factorized by matching the QCD currents and measurement operators onto SCET dijet and measurement operators, which in the formulation of \cite{Freedman:2011kj}, explicitly decouple the \ncoll, \nbcoll, and soft degrees of freedom.  The non-local product of the dijet and measurement operators were then matched onto jet and soft operators that separately describe each of these degrees of freedom.  The approach illustrated here can be applied to other jet observables to calculate subleading corrections.  We are currently exploring subleading corrections to the continuous angularity observables of which thrust is an example.


\acknowledgments{This work was supported by the Natural Sciences and Engineering Research Council of Canada  We would like to thank M. Luke for useful insight throughout this project.}


\begin{appendix}


\bwtxt
\section{Dijet Operators\label{app:operators}}

The LO dijet operator was given in \eqn{O2lo}.  In this section we write the NLO and \nnlo\ dijet operators necessary for calculating the $O(\al_s\tau)$ thrust rate.  They are found by following the SCET formulation of \cite{Freedman:2011kj}.  The NLO corrections to the \ncoll\ sector are described by the operators \cite{Freedman:2011kj}
\ba
	O_2^{(1a_n)}(x)&=&\left[\bar\psi_n(\xnb)P_\nb\Gamma i\Dslash_\perp(\xnb)\frac{\nslash}{2}\wnfun(\xnb,\xnbinf)\right]\left[\ynfun(\xsninf, 0)\ynbfun(0,\xsnbinf)\right]\left[\wnbfun(\xninf,\xn)\psi_\nb(\xn)\right]\nn
	O_2^{(1b_n)}(x)&=&\left[\bar\psi_n(\xnb)\frac{\nbslash}{2}i\overleftarrow{\Dslash}_\perp(\xnb)\Gamma\wnfun(\xnb,\xnbinf)\right]\left[\ynfun(\xsninf, 0)\ynbfun(0,\xsnbinf)\right]\left[\wnbfun(\xninf,\xn)P_\nb\psi_\nb(\xn)\right]\non
\ea
\ba
	O_2^{(1e_{n})}(x)&=&-i\int_0^\infty dt \left[\bar{\psi}_n(\xnb+t\nb)T^d\frac{\fmslash{\nb}}2 \gn_\perp\Gamma \wnfun(\xnb+t\nb, \xnb)P_n\psi_n(\xnb)\wnadj{}^{dc}(\xnb+t\nb,\xnb^{\infty})\right]\nn
		&\times&\left[\ynadj{}^{c\hat b}(\xsninf, 0)\ynbadj{}^{\hat bb}(0, \xsninf)\right]\left[n_\mu\wnbadj{}^{ba}(\xninf,\xn) igG_\nb^{a\mu\nu}(\xn)\right]\\
	O^{(1f_{n})}_2(x)&=&-i\int_0^\infty dt \left[\bar{\psi}_n(\xnb)P_\nb\Gamma\gn_\perp\frac{\fmslash{\nb}}2  \wnfun(\xnb, \xnb+t \nb)T^d\psi_n(\xnb+t\nb)\wnadj{}^{dc}(\xnb+t\nb,\xnb^{\infty})\right]\nn
		&\times&\left[\ynadj{}^{c\hat b}(\xsninf, 0)\ynbadj{}^{\hat bb}(0, \xsninf)\right]\left[n_\mu\wnbadj{}^{ba}(\xninf,\xn) igG_\nb^{a\mu\nu}(\xn)\right]\nonumber.
\ea
The covariant derivatives are defined as $D^\mu=\partial^\mu-igA^\mu$ and the gauge fields are in the same sector as the fields the derivative is acting on.  The $m^{\rm th}$-sector field strength tensor denoted as $G_m^{\mu\nu}$.  For the corrections to the \nbcoll\ sector, the two operators required are \cite{Freedman:2011kj}
\ba\label{O21n}
	O_2^{(1a_\nb)}(x)&=&\left[\bar\psi_n(\xnb)\wnfun(\xnb,\xnbinf)\right]\left[\ynfun(\xsninf, 0)\ynbfun(0,\xsnbinf)\right]\left[\wnbfun(\xninf,\xn)\frac{\nbslash}{2}i\overleftarrow\Dslash_\perp(\xn)\Gamma P_\nb\psi_\nb(\xn)\right]\nn
	O_2^{(1b_\nb)}(x)&=&\left[\bar\psi_n(\xnb)P_\nb\wnfun(\xnb,\xnbinf)\right]\left[\ynfun(\xsninf, 0)\ynbfun(0,\xsnbinf)\right]\left[\wnbfun(\xninf,\xn)\Gamma i\Dslash_\perp(\xn) \frac{\nslash}{2}\psi_\nb(\xn)\right].
\ea
See \cite{Freedman:2011kj} for the Feynman rules and physical pictures of these operators.  The matching coefficients are
\ba
	C_2^{(1a_n)}=C_2^{(1a_\nb)}=-1+O(\al_s),\quad	C_2^{(1b_n)}=C_2^{(1b_\nb)}=1+O(\al_s),\quad	C_2^{(1e_{n})}=-1+O(\al_s), \quad	C_2^{(1f_{n})}=1+O(\al_s).\nn
\ea
The operators describing the \nnlo\ corrections to the \ncoll\ sector are found in the same way as the NLO operators.  The operators are
\ba\label{O22n}
	O_2^{(2a_n)}(x)&=&\left[\bar\psi_n(\xnb)P_\nb\Gamma in\cdot D(\xnb)\wnfun(\xnb,\xnbinf)\right]\left[\ynfun(\xsninf, 0)\ynbfun(0,\xsnbinf)\right]\left[\wnbfun(\xninf,\xn)P_\nb\psi_\nb(\xn)\right]\nn
	O_2^{(2b_n)}(x)&=&-i\int_0^\infty dt\left[\bar\psi_n(\xnb+t\nb)i\overleftarrow\Dslash_\perp(\xnb+t\nb)\frac{\nbslash}{2}\Gamma\frac{\nslash}{2}\wnfun(\xnb+t\nb,\xnb) i\Dslash_\perp(\xnb)\wnfun(\xnb,\xnbinf)\right]\nn
		&\times&\left[\ynfun(\xsninf, 0)\ynbfun(0,\xsnbinf)\right]\left[\wnbfun(\xninf,\xn)\psi_\nb(\xn)\right]\nn
	O_2^{(2A_n)}(x)&=&-i\int_0^\infty dt\,\left[\bar\psi_n(\xnb)P_\nb\wnfun(\xnb,\xnb+t\nb)\Gamma g\si_{\perp\al\beta}G_n^{\al\beta}(\xnb+t\nb)\wnfun(\xnb+t\nb,\xnbinf)\right]\nn
		&\times&\left[\ynfun(\xsninf, 0)\ynbfun(0,\xsnbinf)\right]\left[\wnbfun(\xninf,\xn)P_\nb\psi_\nb(\xn)\right]\\
	O_2^{(2\de_n)}(x)&=&Q\left[\bar\psi_n(\xnb)P_\nb\Gamma\left(n\cdot D+n\cdot\overleftarrow{D}\right)(\xnb)\wnfun(\xnb,\xnbinf)\right]\left[\ynfun(\xsninf, 0)\ynbfun(0,\xsnbinf)\right]\left[x^-\wnbfun(\xninf,\xn)P_\nb\psi_\nb(\xn)\right]\nonumber
\ea
with matching coefficients
\ba
	C_2^{(2a_n)}&=-\frac12+O(\al_s), \qquad &C_2^{(2b_n)}=-1+O(\al_s), \qquad	C_2^{(2A_n)}=\frac12+O(\al_s),\qquad C_2^{(2\de_n)}=1+O(\al_s).
\ea
The $Q$ in $O_2^{(2\de_n)}$ is required because of the definition in \eqn{qcdmatch}.  The operators describing corrections to the \nbcoll\ sector are
\ba
	O_2^{(2a_\nb)}(x)&=&\left[\bar\psi_n(\xnb)P_\nb\wnfun(\xnb,\xnbinf)\right]\left[\ynfun(\xsninf, 0)\ynbfun(0,\xsnbinf)\right]\left[\wnbfun(\xninf,\xn)\Gamma i\nb\cdot D(\xn)P_\nb\psi_\nb(\xn)\right]\nn
	O_2^{(2b_\nb)}(x)&=&-i\int_0^\infty dt\left[\bar\psi_n(\xnb)\wnfun(\xnb,\xnbinf)\right]\left[\ynfun(\xsninf, 0)\ynbfun(0,\xsnbinf)\right]\nn
		&\times&\left[\wnbfun(\xninf,\xn)i\overleftarrow\Dslash_\perp(\xn)\wnbfun(\xn,\xn+tn)\frac{\nbslash}{2}\Gamma\frac{\nslash}{2}i\Dslash(\xn+tn)\psi_\nb(\xn+tn)\right]\nn
	O_2^{(2A_\nb)}(x)&=&-i\int_0^\infty dt\,\left[\bar\psi_n(\xnb)P_\nb\wnfun(\xnb,\xnbinf)\right]\left[\ynfun(\xsninf, 0)\ynbfun(0,\xsnbinf)\right]\nn
		&\times&\left[\wnbfun(\xninf,\xn+tn)gG_\nb^{\al\beta}(\xn+tn)\si_{\perp\al\beta}\Gamma P_\nb \wnfun(\xn+tn,\xn)\psi_\nb(\xn)\right]\\
	O_2^{(2\de_\nb)}(x)&=&Q\left[x^+\bar\psi_n(\xnb)P_\nb\wnfun(\xnb,\xnbinf)\right]\left[\ynfun(\xsninf, 0)\ynbfun(0,\xsnbinf)\right]\left[\wnbfun(\xninf,\xn)\Gamma\left(\nb\cdot D+\nb\cdot\overleftarrow{D}\right)(\xn)P_\nb\psi_\nb(\xn)\right]\nonumber
\ea
with matching coefficients the same as their \ncoll\ counterparts.  The intuitive picture of these operators is similar to the pictures presented in \cite{Freedman:2011kj}.

The NLO corrections to the soft sector are described by the operators \cite{Freedman:2011kj}
\ba 
	O_2^{(1c_{ns})}(x)&=&-i\int_0^{\infty} dt \left[\bar\psi_n(\xnb)P_{\bar n} \Gamma i\overleftarrow{D}^\mu_{\perp}(\xnb) \wnfun(\xnb, \xnbinf)\right]\nn
		&\times&\left[\ynfun(\xsninf,t n) i\overleftarrow{D}_{\perp\mu} (t n)\ynfun(t n, 0)  \ynbfun(0, \xsnbinf)\right]\left[ \wnbfun(\xninf, \xn)P_\nb\psi_\nb(\xn)\right]\nn
	O_2^{(1c_{\nb s})}(x)&=&-i\int_0^{\infty} dt \left[\bar\psi_n(\xnb)P_{\bar n} \Gamma\wnfun(\xnb, \xnbinf)\right]\left[\ynfun(\xsninf,0)\ynbfun(0,t\nb)i{D}_{\perp\mu} (t \nb)  \ynbfun(t\nb, \xsnbinf)\right]\nn
		&\times&\left[ \wnbfun(\xninf, \xn) iD^\mu_{\perp}(\xn)P_\nb\psi_\nb(\xn)\right]\\
	O^{(1\delta_{s})}_2(x)&=&Q\left[x_{\perp\mu}\bar\psi_n(\xnb)P_{\bar n} \Gamma\wnfun(\xnb,\xnbinf)\right]\left[\ynfun(\xsninf, 0) (D_\perp^\mu+\overleftarrow{D}_\perp^\mu 
)\ynbfun(0, \xsnbinf)\right]\left[ \wnbfun(\xninf, \xn)P_\nb\psi_\nb(\xn)\right]\nn
	O^{(1d_{ns})}_2(x)&=&-i\int_0^\infty dt \left[\nb_\mu igG_n^{a\mu\nu}(\xnb)\wnadj{}^{ab}(\xnb,\xnbinf)\right]\nn
	&&\times\left[\ynadj{}^{bc}(\xsninf,tn)\bar{\psi}_{s}(tn)T^c\ynfun(tn,0) \frac{\fmslash{n}}{2}\gn_\perp\Gamma \ynbfun(0,\xsnbinf)\right]\left[\wnbfun(\xninf, \xn)P_\nb\psi_\nb(\xn)\right]\nn
	O^{(1d_{\nb s})}_2(x)&=&-i\int_0^\infty dt \left[\bar\psi_n(\xnb)P_\nb\wnfun(\xnb,\xnbinf)\right]\left[\ynfun(\xsninf,0)\Gamma\gamma_{\nu\perp}\frac{\nbslash}{2}T^c\ynbfun(0,t\nb) \psi_s(t\nb)\ynbadj{}^{cb}(t\nb,\xsnbinf)\right]\nn
	&&\times\left[\wnbadj{}^{ba}(\xninf,\xn)n_\mu igG_\nb^{a\mu\nu}(\xn)\right],\nonumber
\ea
with matching coefficients 
\ba
	C_2^{(1c_{ns})}=C_2^{(1c_{\nb s})}=-2+O(\al_s), \quad C_2^{(1d_{ns})}=C_2^{(1d_{\nb s})}=1+O(\al_s), \quad C_2^{(1\de_s)}=1+O(\al_s).
\ea
The Feynman rules of these operators are given in \cite{Freedman:2011kj}.  The \nnlo\ corrections for soft gluon emission are
\ba
	O_2^{(2A_{ns})}(x)&=&-i\int_0^\infty dt\left[\bar\psi_n(\xnb)P_\nb\wnfun(\xnb,\xnbinf)\right]\left[\ynfun(\xsninf, t n) g \si_{\mu\nu}G_s^{\mu\nu}(tn)\ynfun(tn,0)\Gamma\ynbfun(0,\xsnbinf)\right]\nn
		&&\times\left[\wnbfun(\xninf,\xn)P_\nb\psi_\nb(\xn)\right]\nn
	O_2^{(2A_{\nb s})}(x)&=&-i\int_0^\infty dt\left[\bar\psi_n(\xnb)P_\nb\wnfun(\xnb,\xnbinf)\right]\left[\ynfun(\xsninf, 0)\Gamma\ynbfun(0,t\nb) g \si_{\mu\nu}G_s^{\mu\nu}(t\nb)\ynbfun(t\nb,\xsnbinf)\right]\nn
		&&\times\left[\wnbfun(\xninf,\xn)P_\nb\psi_\nb(\xn)\right]
\ea
with matching coefficients
\ba
	C_2^{(2A_{ns})}=C_2^{(2A_{\nb s})}=\frac12+O(\al_s).
\ea
The corrections from the expansion of the momentum conserving delta function are described by
\ba
	O_2^{(2\de_{s+})}(x)&=&Q\left[x^+\bar\psi_n(\xnb)P_\nb\Gamma\wnfun(\xnb,\xnbinf)\right]\left[\ynfun(\xsninf, 0)\left(\nb\cdot D+\nb\cdot\overleftarrow{D}\right)\ynbfun(0,\xsnbinf)\right]\left[\wnbfun(\xninf,\xn)P_\nb\psi_\nb(\xn)\right]\nn
	O_2^{(2\de_{s-})}(x)&=&Q\left[\bar\psi_n(\xnb)P_\nb\Gamma\wnfun(\xnb,\xnbinf)\right]\left[\ynfun(\xsninf, 0)\left(n\cdot D+n\cdot\overleftarrow{D}\right)\ynbfun(0,\xsnbinf)\right]\left[x^-\wnbfun(\xninf,\xn)P_\nb\psi_\nb(\xn)\right]\nn
	O_2^{(2\de_{s\perp})}(x)&=&\frac Q2\left[x_{\perp\al}x_{\perp\beta}\bar\psi_n(\xnb)P_\nb\Gamma\wnfun(\xnb,\xnbinf)\right]\left[\ynfun(\xsninf, 0)\left(D_\perp^\al D_\perp^\beta+\overleftarrow{D}_\perp^\beta\overleftarrow{D}_\perp^\al \right)\ynbfun(0,\xsnbinf)\right]\nn
		&&\times\left[\wnbfun(\xninf,\xn)P_\nb\psi_\nb(\xn)\right]\nn
	O_2^{(2\de c_{ sn})}(x)&=&-i\int_0^{\infty} dt \left[x_{\perp\mu}\bar\psi_n(\xnb)P_{\bar n} \Gamma i\overleftarrow{D}^\nu_{\perp}(\xnb) \wnfun(\xnb, \xnbinf)\right]\\
		&\times&\left[\ynfun(\xsninf,t n) i\overleftarrow{D}_{\perp\nu} (t n)\ynfun(t n, 0) ( D_\perp^\mu+\overleftarrow{D}_\perp^\mu)(0)  \ynbfun(0, \xsnbinf)\right]\left[ \wnbfun(\xninf, \xn)P_\nb\psi_\nb(\xn)\right]\nn
	O_2^{(2\de c_{ s\nb})}(x)&=&-i\int_0^{\infty} dt \left[x_{\perp\mu}\bar\psi_n(\xnb)P_{\bar n} \wnfun(\xnb, \xnbinf)\right]\nn
		&\times&\left[\ynfun(\xsninf,0)( D_\perp^\mu+\overleftarrow{D}_\perp^\mu)(0) \ynbfun(0,t\nb) iD_{\perp\nu} (t \nb)\ynbfun(t\nb, \xsnbinf)\right]\left[ \wnbfun(\xninf, \xn)iD_{\nu\perp}(\xn)\Gamma P_\nb\psi_\nb(\xn)\right]\nonumber
\ea
with matching coefficients
\ba
	C_2^{(2\de_{s+})}=C_2^{(2\de_{s-})}=1+O(\al_s),\quad C_2^{(2\de_{s\perp})}=1+O(\al_s), \quad C_2^{(2\de c_{sn})}=C_2^{(2\de c_{s\nb})}=-2+O(\al_s).
\ea
We do not require the \nnlo\ corrections from soft quark emission for an \nnlo\ calculation.

As expected, the subleading operators can be written as separate \ncoll, \nbcoll, and soft pieces.  The Feynman rules for the LO and NLO operators at $O(\al_s)$ were shown in \cite{Freedman:2011kj}.  For the sake of brevity, we do not show the \nnlo\ Feynman rules here.   We are only concerned with vector currents in the above calculations, so $\Gamma=\ga^\mu$.


\section{Jet and Soft Operators\label{app:jjs}}

The jet and soft operators are found by doing the matching in \eqn{ope}.  It is convenient to use the basis
\ba\label{fierz}
	\left\{1,\ga_\perp^\al,\ga_5,\si_\perp^{\al\beta},\left[\frac{\nslash}2,\frac{\nbslash}2\right], \ga_\perp^\al\ga_5,\frac{\nslash}2,\frac{\nbslash}2,\frac{\nslash}{2}\ga_\perp^\al,\frac{\nbslash}2\ga_\perp^\al,\frac{\nslash}{2}\ga_5,\frac{\nbslash}{2}\ga_5 \right\}
\ea
for the Dirac matrices.  We have defined $\ga_\perp^\mu=g_{\perp}^{\mu\nu}\ga_\nu$ and $\si_\perp^{\mu\nu}=\frac i2[\ga_\perp^\mu,\ga_\perp^\nu]$ where $g_\perp^{\mu\nu}=g^{\mu\nu}-(n^\mu\nb^\nu+n^\nu\nb^\mu)/2$.  For use later, we aslo define $\ep_\perp^{\al\beta}=n_\mu\nb_\nu\ep^{\al\beta\mu\nu}$. The jet and soft operators will be parity even scalars due to only considering vector currents.

\begin{figure*}[t]
	\centering
	\subfloat[$\vac{J^{(0_{\bf 8})}}$\label{fig:0dns}]{\includegraphics[width=0.15\textwidth]{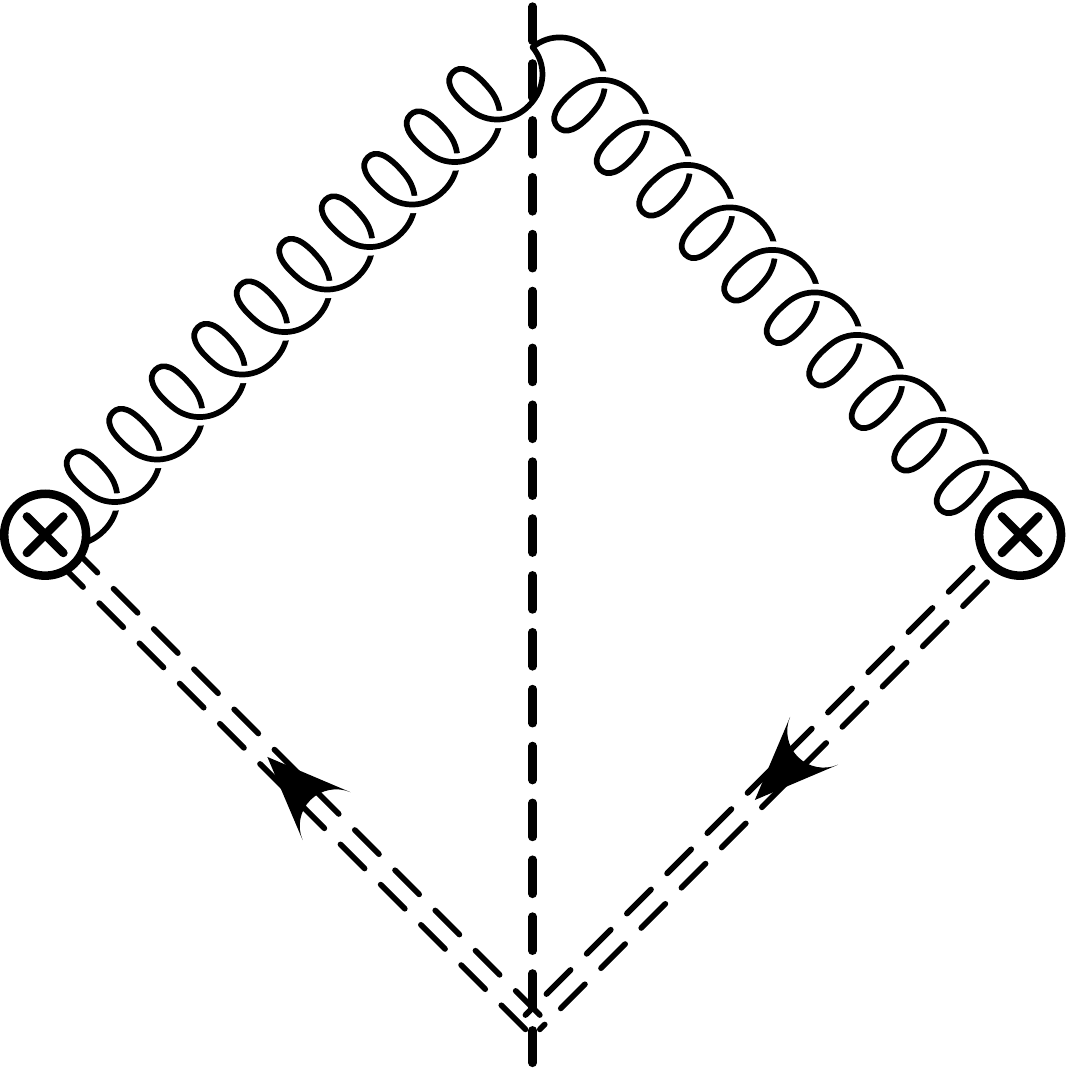}}\hspace{0.1\textwidth}
	\subfloat[$\vac{J^{(2{e})}}$\label{fig:2e}]{\bop[width=0.15\textwidth]{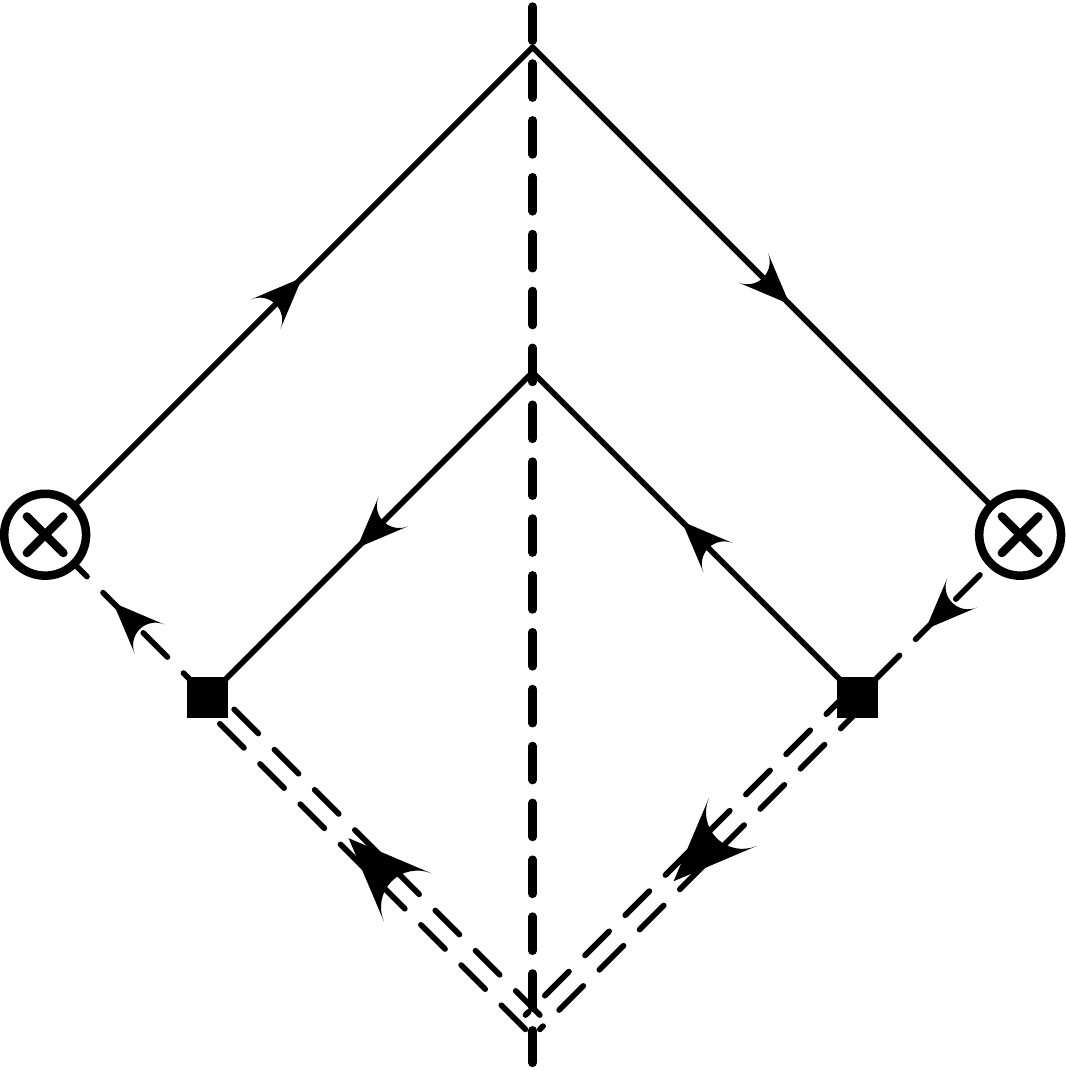}
		\put(85,30){\footnotesize$s\nb$}
		\put(-25,30){\footnotesize$\xnb+t\nb$}
		\eop} \hspace{0.1\textwidth}
	\subfloat[$\vac{S^{(2{d_{ns}})}}$\label{fig:S2dns}]{\bop[width=0.15\textwidth]{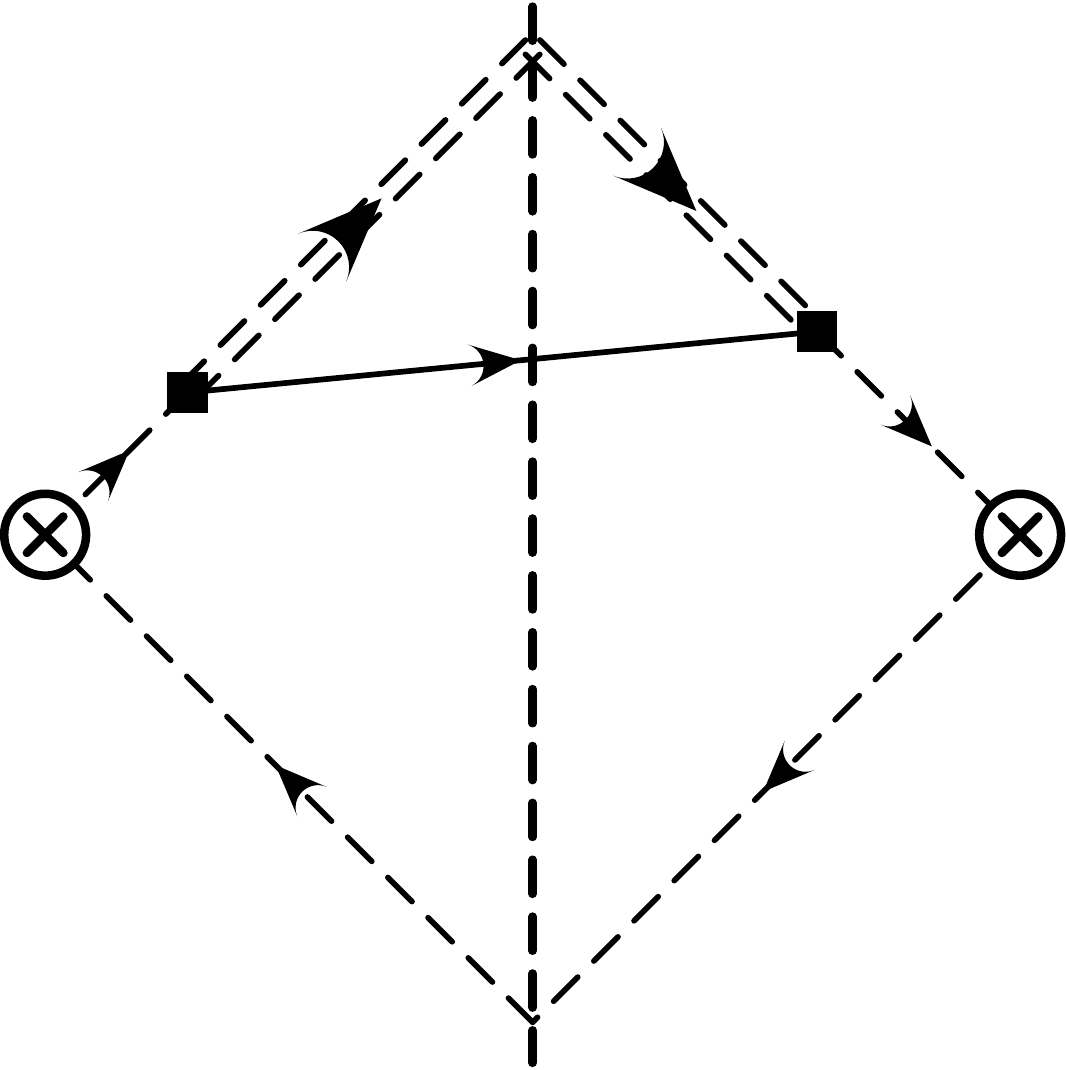}
		\put(2,60){\footnotesize$tn$}
		\put(80,67){\footnotesize$sn$}
		\eop}\hspace{0.1\textwidth}
	\subfloat[$\vac{S^{(0_{\bf 8})}}$\label{fig:S0e}]{\includegraphics[width=0.15\textwidth]{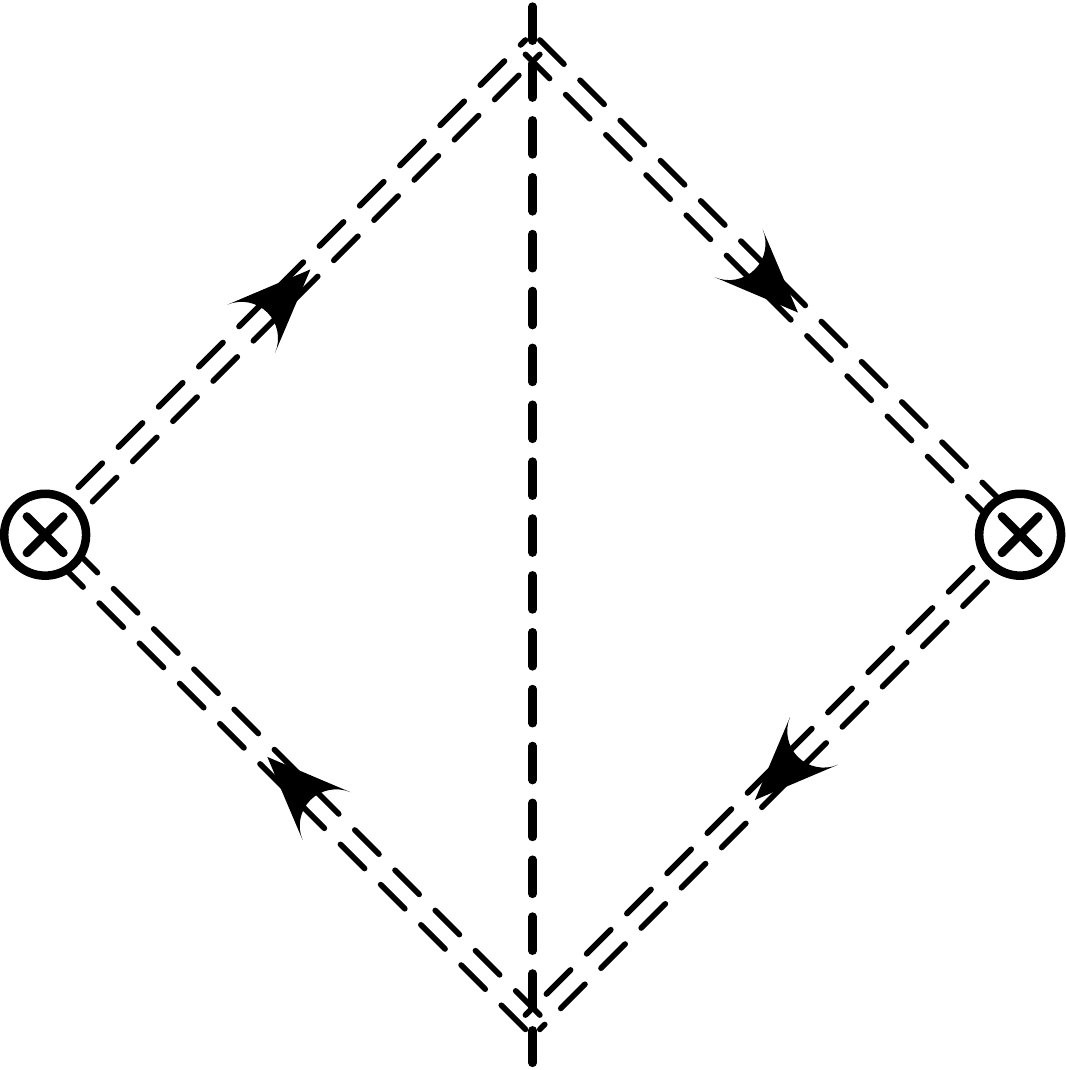}}
	\fcaption{Diagrams for the operators describing a soft quark and the quark and anti-quark in the same sector.  The double dashed lines represent Wilson lines in the adjoint representation.   Reverse the fermion arrows for $\vac{J^{(2{f})}}$.  The filled boxes highlight a Wilson line changing representation.\label{fig:JJSnlo}}
\end{figure*}

The leading order jet operators required at $O(\al_s)$ are
\ba
	J^{(0)}(\mu,\tau)&=&\frac{1}{N_C}\Tr\int \dn \bar\psi_n(0)\wnfun(0,\xsnbinf)\frac{\nbslash}{2}\hat\M_n^{(0)}(\tau)\wnfun(\xnbinf,\xnb)\psi_n(\xnb)\nn
	J^{(0_{\bf 8})}(\mu,\tau)&=&\frac{g_\perp^{\al\beta}\nb_\mu\nb_\nu}{QN_C}\int\dn G_n^{\mu\al a}(0)\wnadj^{ab}(0,\xsninf)\hat\M_n^{(0)}(\tau)\wnadj^{b\bar a}(\xnbinf,\xnb)G_n^{\nu\beta\bar a}(\xnb)\nn
	J^{(0\de_\nb)}(\mu,\tau)&=&\frac{Q}{N_C}\Tr\int \dn x^+\bar\psi_n(0)\wnfun(0,\xsnbinf)\frac{\nbslash}{2}\hat\M_n^{(0)}(\tau)\wnfun(\xnbinf,\xnb)\psi_n(\xnb)\nn
	J^{(0c_{ns}\de_s)}(\mu,\tau)&=&\frac {1}{N_C}\Tr\int \dn x_{\perp\al}\bar\psi_n(0) \overleftarrow D^\al_\perp(0)\wnfun(0,\xsninf)\frac{\nbslash}{2}\hat\M^{(0)}(\tau)\wnfun(\xnbinf,\xnb)\psi_n(\xnb)\nonumber.
\ea
Although $J^{(0c_{ns}\de_s)}$ has a $D_\perp$ suppression, there is an enhancement by an explicit $x_\perp$ so the operator is LO.  The vacuum expectation value of the operator $J^{(0_{\bf 8})}$ is shown in \fig{0dns}.  The other operators have the same diagrams as \fig{LO} with the appropriate derivative insertions.   The $Q$'s are introduced by dimensional analysis.  There are also the enhanced jet operators
\ba
	J^{(-2\de_s)}(\mu,\tau)&=&\frac{Q^2}{N_C}\Tr\int \dn (x_\perp)^2\bar\psi_n(0)\wnfun(0,\xsnbinf)\frac{\nbslash}{2}\hat\M_n^{(0)}(\tau)\wnfun(\xnbinf,\xnb)\psi_n(\xnb)\\
	J^{(-2\de_s M_n)}(\mu,\tau)&=&\frac{1}{N_C}\Tr\int \dn x_{\perp\al} \bar\psi_n(0) \wnfun(0,\xsninf)\frac{\nbslash}2\frac{\overleftarrow{\partial}_\perp^\al}{Q}\frac{\partial}{\partial\tau}\hat\M_n^{(0)}(\tau)\wnfun(\xnbinf,\xnb)\psi_n(\xnb)\nn
	J^{(-2M_{2n_b})}(\mu,\tau)&=&\frac{1}{N_C}\Tr\int \dn  \bar\psi_n(0) \wnfun(0,\xsninf)\frac{\nbslash}2\frac{\nb\cdot\overleftarrow\partial}{Q}\frac{\partial}{\partial\tau}\hat\M_n^{(0)}(\tau)\wnfun(\xnbinf,\xnb)\psi_n(\xnb)\nonumber
\ea
where the derivatives act at the cut.  These operators are always associated with $O(\lambda^4)$ soft operators so do not give an enhancement to the thrust rate.  We have neglected the contribution from $\M^{(-2n_a)}$ because it vanishes at $O(\al_s)$.    The required \nnlo\ jet operators are
\ba
	J^{(2ab_n)}(\mu,\tau)&=&\frac{1}{Q^2N_C}\Tr\int\dn \bar\psi_n(0)D_\perp^\al(0)\wnfun(0,\xsninf)\frac{\nbslash}{2}\hat\M^{(0)}_n(\tau)\wnfun(\xnbinf,\xnb) D_{\perp\al}(\xnb)\psi_n(\xnb)\nn
	J^{(2e)}(\mu,\tau)&=&\frac{g^2}{QN_C}\Tr\int_0^\infty dsdt\int\dn \wnadj^{ab}(\xnbinf,\xnb+t\nb)\bar\psi_n(\xnb)\wnfun(\xnb,\xnb+t\nb)\nn
		&&\times\gamma_\perp^\mu\gamma_{\perp\al}\frac{\nbslash}{2}T^b\psi_n(\xnb+t\nb)\hat\M_n^{(0)}(\tau)\bar\psi_n(s\nb)T^{\bar b}\frac{\nbslash}{2}\gamma_{\perp\al}\gamma_{\perp\mu}\wnfun(s\nb,0)\psi_n(0)\wnadj^{\bar ba}(s\nb,\xsninf)\nn
	J^{(2f)}(\mu,\tau)&=&\frac{g^2}{QN_C}\Tr\int_0^\infty dsdt\int\dn \wnadj^{ab}(\xnbinf,\xnb+t\nb)\bar\psi_n(\xnb+t\nb)T^b\wnfun(\xnb+t\nb,\xnb)\nn
		&&\times\frac{\nbslash}{2}\gamma_\perp^\al\gamma_\perp^\mu \psi_n(\xnb)\hat\M_n^{(0)}\bar\psi_n(0)\gamma_{\perp\mu}\gamma_{\perp\al}\frac{\nbslash}{2}\wnfun(0,s\nb)T^{\bar b}\psi_n(s\nb)\wnadj^{\bar ba}(t\nb,\xsninf)\\
	J^{(2a_n)}(\mu,\tau)&=&\frac{1}{QN_C}\Tr\int\dn\bar\psi_n(0)in\cdot D(0)\wnfun(0,\xsninf)\frac{\nbslash}{2}\hat\M^{(0)}_n(\tau)\wnfun(\xnbinf,\xnb)\psi_n(\xnb)\nn
	J^{(2A_n)}(\mu,\tau)&=&\frac{1}{QN_C}\Tr\int_0^\infty dt\int\dn \bar\psi_n(0)\wnfun(0,t\nb)g\ep_\perp^{\al\beta}G_n^{\al\beta}(t\nb)\wnfun(t\nb,,\xsninf)\nn
		&&\times\frac{\nbslash}{2}\gamma_5\hat\M_n^{(0)}(\tau)\wnfun(\xnbinf,\xnb)\psi_n(\xnb)\nn
	J^{(2\de_n)}(\mu,\tau)&=&\frac{1}{N_C}\Tr\int \dn \bar\psi_n(0)\wnfun(0,\xsnbinf)\frac{\nbslash}{2}\hat\M_n^{(0)}(\tau)\wnfun(\xnbinf,\xnb)\left(in\cdot \overleftarrow D+in\cdot D\right)(\xnb)\psi_n(\xnb)\nonumber
\ea
and their hermitian conjugates.  The vacuum expectation value of $J^{(2ab_n)}$ was pictured in \fig{j2abn}.  Although $J^{(2e)}$ and $J^{(2f)}$ look complicated, they have simple diagrams shown in \fig{2e}.   The $\gamma_5$ in $J^{(2A_n)}$ is necessary because $\ep_\perp^{\al\beta}$ is parity odd.  The vacuum expectation value of these operators are shown in \tab{vevJ}.

\begin{table*}
	{\renewcommand{\arraystretch}{1.25}
	\subfloat[\label{tab:vevJ}]{
		$\begin{array}{|c|c|}
			\hline
			l 			& \vac{J^{(l)}(\mu,\tau)} \\
			\hline\hline
			 0 			& \de(\tau) \\
			 0_{\bf 8} 		& -\de(\tau) \\
			 0\de_n 		& -2\de(\tau) \\
			 0c_{ns}\de_s 	& -2\de(\tau) \\ 
			\hline
			 -2\de_s		& 2\de'(\tau)\\
			 -2 \de_s M_n 	& \de'(\tau) \\
			 -2M_{n_b} 	& \de'(\tau) \\ 
			\hline
	 		 2ab_n 		& \alsb /2 \\
			 2a_n 		& \alsb(-2/\ep-2\Ln-9/2)\\
			 2A_n 		& \alsb \\
			 2\de_n 		& \alsb (-2/\ep -2\Ln-3/2)\\
			 2e 			& \alsb(-2/\ep-2\Ln+2)\\
			 2f 			& \alsb(-2/\ep-2\Ln+2)\\ 
			\hline
		\end{array}$
		}
\hspace{0.05\textwidth}
	\subfloat[\label{tab:vevJb}]{
		$\begin{array}{|c|c|}
			 \hline
			 m 			& \vac{\bar J^{(m)}(\mu,\tau)} \\
			\hline\hline
			  0 			& \de(\tau)\\ 
			  0_{\bf 8} 	& -\de(\tau)\\
			  0_{M1} 		& \de(\tau)\\
			  0_{M2} 		& \de(\tau)\\
			  0\de_n 		& -2\de(\tau)\\
			\hline
			  -2M_{\nb_b} 	& \de'(\tau)\\
			\hline
			  2ab_\nb 		& \alsb/2 \\
			  2a_\nb 		& \alsb(-2/\ep-2\Ln-9/2)\\
			  2A_\nb 		& \alsb \\
			  2\de_\nb 	& \alsb(-2/\ep-2\Ln-3/2)\\
			\hline
		\end{array}$
		}
\hspace{0.05\textwidth}
	\subfloat[\label{tab:vevS}]{
		$\begin{array}{|c|c|}
			\hline
			 n 			& \vac{S^{(n)}(\mu,\tau)} \\ 
			 \hline \hline
			  0 			& \de(\tau) \\
			  0_{\bf 8} 	& \de(\tau) \\
			  \hline
			  2c_{ns}\de_s 	& \alsb(1/\ep+\Ls-1) \\
			  2\de c_{ns} 	& \alsb(1/\ep+\Ls -1)\\
			  2d_{ns} 		& \alsb(-1/\ep-\Ls+1)\\
			  2d_{\nb s} 	& \alsb(-1/\ep-\Ls+1)\\
			  2A_{ns} 		& \alsb(2/\ep+2\Ls-2)\\
			  2\de_{sn} 	& \alsb(2/\ep+2\Ls-2)\\
			  2M_{s_1}	& -4\alsb\\
			  2M_{s_{+t}} 	& -4\alsb\\
			  \hline
			  4\de_{s\perp} & -4\alsb\tau \\
			  4\de_sM_n 	& 4\alsb\tau\\
			  4M_{n_b} 	& 2\alsb\tau \\
			  4M_{\nb_b} 	& -2\alsb\tau\\
			  \hline
		\end{array}$
		}
}
	\fcaption{\label{tab:vev}  Relevant vacuum expectation values of the jet and soft operators for the $O(\al_s\tau)$ thrust rate.  The operators distinguished by primes give the same values.  }
\end{table*}

The $\bar J$ operators are similar to the $J$  operators.  The leading order operators are
\ba
	\bar J^{(0)}(\mu,\tau)&=&\frac{1}{N_C}\Tr\int \dnb\bar\psi_\nb(\xn)\wnbfun(\xn,\xninf)\frac{\nslash}{2}\hat\M_\nb^{(0)}(\tau)\wnbfun(\xsninf,0)\psi_\nb(0)\nn
	\bar J^{(0_{\bf 8})}(\mu,\tau)&=&\frac{g_\perp^{\al\beta}n_\mu n_\nu}{Q}\int\dnb G_\nb^{\mu\al a}(\xn)\wnbadj^{ab}(\xn,\xninf)\hat\M_\nb^{(0)}(\tau)\wnbadj^{b\bar a}(\xsnbinf,0)G_\nb^{\nu\beta\bar a}(0)\non
\ea
\ba
	\bar J^{(0_{M1})}(\mu,\tau)&=&\frac{iQ}{N_C}\Tr\int_0^\infty dt\int \dnb\bar\psi_\nb(\xn)\wnbfun(\xn,\xninf)\frac{\nslash}{2}\hat\M_\nb^{(0)}(\tau)\wnbfun(\xsninf,tn)\psi_\nb(tn)\nn
	\bar J^{(0_{M2})}(\mu,\tau)&=&\frac{iQ^2}{N_C}\Tr\int_0^\infty dt\int_0^\infty ds\int \dnb \bar\psi_\nb(\xn)\wnbfun(\xn,\xninf)\frac{\nslash}{2}\hat\M_\nb^{(0)}(\tau)\wnbfun(\xsninf,(t+s)n)\psi_\nb((t+s)n)\nn
	\bar J^{(0\de_n)}(\mu,\tau)&=&\frac{iQ}{N_C}\Tr\int \dnb x^-\bar\psi_\nb(\xn)\wnbfun(\xn,\xninf)\frac{\nslash}{2}\hat\M_\nb^{(0)}(\tau)\wnbfun(\xsninf,0)\psi_\nb(0)\nonumber.
\ea
There is only one enhanced operator that needs to be considered
\ba
	\bar J^{(-2M_{\nb_b})}(\mu,\tau)&=&\frac{1}{N_C}\Tr\int \dnb\bar\psi_\nb(\xn)\wnbfun(\xn,\xninf)\frac{\nslash}{2}\frac{in\cdot\overleftarrow\partial}{Q}\frac{\partial}{\partial\tau}\hat\M_\nb^{(-2\nb_b)}(\tau)\wnbfun(\xsninf,0)\psi_\nb(0).\nn
\ea
The \nnlo\ operators are
\ba
	\bar J^{(2ab_\nb)}(\mu,\tau)&=&\frac{1}{QN_C}\Tr\int\dnb \bar\psi_\nb(\xn)iD_\perp^\al\wnbfun(\xn,\xsninf)\frac{\nslash}{2}\hat\M^{(0)}_\nb(\tau)\wnbfun(\xninf,\xn)i D_{\perp\al}(\xn)\psi_\nb(0)\nn
	\bar J^{(2a_\nb)}(\mu,\tau)&=&\frac{1}{QN_C}\Tr\int\dnb\bar\psi_\nb(\xn)\wnbfun(\xn,\xsninf)\frac{\nslash}{2}\hat\M^{(0)}_\nb(\tau)\wnbfun(\xninf,0)i\nb\cdot \overleftarrow D(0)\psi_\nb(\xn)\\
	\bar J^{(2A_\nb)}(\mu,\tau)&=&\frac{1}{QN_C}\Tr\int_0^\infty dt\int\dn \bar\psi_n(\xn)\wnbfun(\xn,\xninf)\gamma_5\frac{\nslash}{2}\hat\M_\nb^{(0)}(\tau)\nn
		&&\times\wnbfun(\xnbinf,tn)g\ep_\perp^{\al\beta}G_\nb^{\al\beta}(tn)\wnfun(tn,0)\psi_\nb(0)\nn
	\bar	J^{(2\de_\nb)}(\mu,\tau)&=&\frac{1}{N_C}\Tr\int \dnb \bar\psi_\nb(\xn)\left(i\nb\cdot \overleftarrow D+i\nb\cdot D\right)(\xn)\wnbfun(\xn,\xsninf)\frac{\nslash}{2}\hat\M_\nb^{(0)}(\tau)\wnbfun(\xninf,0)\psi_\nb(0)\nonumber
\ea
The diagrams are similar to those found for the $J$ operators.   For example, $J^{(0_{\bf 8})}$ is the horizontal reflection of \fig{0dns}.  The vacuum expectation values of these operators are shown in \tab{vevJb}.  The difference in the required $\bar J$ operators compared to the $J$ operators is because we have chosen the $\vec n$ axis to be anti-parallel to the \nbcoll\ sector.  

The leading order soft operators are
\ba
	S^{(0)}(\mu,\tau)&=&\frac{1}{N_C}\Tr\, \ynbfun(\xsnbinf,0)\ynfun(0,\xsninf)\hat\M_s^{(0)}(\tau)\ynfun(\xsninf,0)\ynbfun(0,\xsnbinf)\nn
	S^{(0_{\bf 8})}(\mu,\tau)&=&\frac{1}{N_C^2}\Tr\, \ynbadj{}^{ab}(\xsnbinf,0)\ynadj{}^{bc}(0,\xsninf)\hat\M_s^{(0)}(\tau)\ynadj{}^{cd}(\xsninf,0)\ynbadj{}^{da}(0,\xsnbinf)
\ea
which only differ in the representation of the Wilson lines.  The vacuum expectation value of $S^{(0_{\bf 8})}$ is pictured in \fig{S0e}.  The \nnlo\ operators are
\ba
	S^{(2c_{ns}\de_s)}(\mu,\tau)&=&\frac{i}{QN_C}\Tr\int_0^\infty dt \ynbfun(\xsnbinf,0)\left(i\overleftarrow D_\perp^\al+iD_\perp^\al\right)(0)\ynfun(0,\xsninf)\hat\M_s^{(0)}(\tau)\nn
		&&\times\ynfun(\xsninf,tn)i\overleftarrow D_\perp^\al(tn)\ynfun(tn,0)\ynbfun(0,\xsnbinf)\nn
	S^{(2\de c_{ns})}(\mu,\tau)&=&\frac{}{QN_C}\Tr\int_0^\infty dt \ynbfun(\xsnbinf,0)\left(i\overleftarrow D_\perp^\al+iD_\perp^\al\right)(0)\ynfun(0,tn)D_{\perp\al}(tn)\ynfun(tn, \xsninf)\hat\M_s^{(0)}(\tau)\nn
		&&\times\ynfun(\xsninf,0)\ynbfun(0,\xsnbinf)\nn
	S^{(2d_{n s})}(\mu,\tau)&=&\frac{g^2}{QN_C^2}\Tr\int_0^\infty dsdt\,\ynadj{}^{ab}(\xsninf,sn)\bar\psi_s(sn)T^b\ynfun(sn,0)\ynbfun(0,\xsnbinf)\frac{\nslash}{2}\hat\M_s^{(0)}(\tau)\nn
		&&\times\ynbfun(\xsnbinf,0)\ynfun(0,tn)T^c\psi_s(tn)\ynadj{}^{ca}(tn,\xsninf)\nn
	S^{(2d_{\nb s})}(\mu,\tau)&=&\frac{g^2}{QN_C^2}\Tr\int_0^\infty dsdt\, \ynbadj{}^{ab}(\xsnbinf,s\nb)\bar\psi_s(s\nb)T^b\ynbfun(s \nb,0)\ynfun(0,\xsninf)\frac{\nbslash}{2}\hat\M_s^{(0)}(\tau)\nn
		&&\times\ynfun(\xsninf,0)\ynbfun(0,t\nb)T^c\psi_s(t\nb)\ynbadj{}^{ca}(t\nb,\xsnbinf)\\
	S^{(2A_{ns})}(\mu,\tau)&=&\frac{gn_\al\nb_\beta}{QN_C}\Tr\int_0^\infty dt\, \ynbfun(\xsnbinf,0)\ynfun(0,tn)G_s^{\al\beta}(tn)\ynfun(tn,\xsninf)\hat\M_s^{(0)}(\tau)\ynfun(\xsninf,0)\ynbfun(0,\xsnbinf)\non
\ea
\ba
	S^{(2A_{\nb s})}(\mu,\tau)&=&\frac{gn_\al\nb_\beta}{QN_C}\Tr\int_0^\infty dt\, \ynbfun(\xsnbinf,t\nb)G_s^{\al\beta}(t\nb)\ynbfun(t\nb,0)\ynfun(0,\xsninf)\hat\M_s^{(0)}(\tau)\ynfun(\xsninf,0)\ynbfun(0,\xsnbinf)\nn
	S^{(2\de_{s\nb})}(\mu,\tau)&=&\frac{1}{QN_C}\Tr\, \ynbfun(\xsnbinf,0)\left(i\nb\cdot\overleftarrow D+i\nb\cdot D\right)(0)\ynfun(0,\xsninf)\hat\M_s^{(0)}(\tau)\ynfun(\xsninf,0)\ynbfun(0,\xsnbinf)\nn
	S^{(2\de_{sn})}(\mu,\tau)&=&\frac{1}{QN_C}\Tr\, \ynbfun(\xsnbinf,0)\left(in\cdot\overleftarrow D+in\cdot D\right)(0)\ynfun(0,\xsninf)\hat\M_s^{(0)}(\tau)\ynfun(\xsninf,0)\ynbfun(0,\xsnbinf)\nonumber
\ea
There are also \nnlo\ operators involving the measurement function expansion that we write as
\ba
	S^{(2M_i)}(\mu,\tau)&=&\frac{1}{N_C}\Tr\, \ynbfun(\xsnbinf,0)\ynfun(0,\xsninf)\hat\M_s^{(2i)}(\tau)\ynfun(\xsninf,0)\ynbfun(0,\xsnbinf)
\ea
where
\ba
	\M_s^{(s_1)}(\tau,\{k\})&=&\frac{2(\kAp)^2}{Q^2}\frac{\partial}{\partial\tau}\M_s^{(0)}(\tau,\{k\})\nn
	\M_s^{(s_{+t})}(\tau,\{k\})&=&\frac{p_\nb^+\nb\cdot\kAnlo}{Q^2}\frac{\partial}{\partial\tau}\M_s^{(0)}(\tau,\{k\})
\ea
where the $p_\nb^+$ cancels in the definition of $\kAnlo$ in \eqn{kAnlo}.  The contribution from the other measurement operators in \eqn{MSnlo} vanish at $O(\al_s)$.

There are also N$^4$LO operators that contribute to the \nnlo\ rate due to convoluting with enhanced jet operators.  These operators are
\ba
	S^{(4\de_{s\perp})}(\mu,\tau)&=&\frac{1}{N_C}\Tr\, \ynbfun(\xsnbinf,0)\left(\overleftarrow D_\perp^\al\overleftarrow D_{\perp\al}+D_\perp^\al D_{\perp\al}\right)\ynfun(0,\xsninf)\hat\M_s^{(0)}(\tau)\ynfun(\xsninf,0)\ynbfun(0,\xsnbinf)\nn
	S^{(4\de_sM_n)}(\mu,\tau)&=&\frac{1}{QN_C}\Tr\,\ynfun(\xsninf, 0)  (iD_\perp^\al+i\overleftarrow{D}_\perp^\al)\ynbfun(0, \xsnbinf)\widehat\M_s^{(2n)\al}(\tau)\ynbfun(\xsnbinf,0)\ynfun(0,\xsninf)\\
	S^{(4 M_i)}(\mu,\tau)&=&\frac{1}{N_C}\Tr\, \ynbfun(\xsnbinf,0)\ynfun(0,\xsninf)\hat\M_s^{(4i)}(\tau)\ynfun(\xsninf,0)\ynbfun(0,\xsnbinf)\nonumber
\ea
where
\ba
	\M_s^{(4n_b)}(\tau,\{k\})&=&\frac{-(\kAp)^2}{Q^2}\M_s^{(0)}(\tau,\{k\})\nn
	\M_s^{(4\nb_b)}(\tau,\{k\})&=&\frac{(\kAp)^2}{Q^2}\M_s^{(0)}(\tau,\{k\}).
\ea
The vacuum expectation value of the soft operators are shown in \tab{vevS}.

\begin{table*}
	\[{\renewcommand{\arraystretch}{1.25}
\begin{array}{c | c ||c||c||c}
	 (i,j,k) 		&	(l,m,n) 			&   C_{1}^{(i,j,k)} & H_{1}^{(i,j,k)}	& R(\tau) \\ \hline \hline
	 (1a_n,1b_n,0)	&	 (2ab_n,0,0) 		& -1  & +1		 	& \alsb\tau/2\\
	 (1b_n,1a_n,0) &	 (2ab_n^\dagger,0,0)& -1 & +1 		& \alsb\tau/2\\
	 (2a_n,0,0)	&	(2a_n,0,0)	  	&  1	& -1/2 		& \alsb\tau(-1/\ep-\Ln-9/4) \\
	 (0,2a_n,0)	&	(2a_n^\dagger,0,0)	&  1	& -1/2 		& \alsb\tau(-1/\ep-\Ln-9/4) \\
	 (2A_n,0,0)	&	(2A_n,0,0)	 	 & -1/2	& -1/4 	& -\alsb\tau/4 \\
	 (0,2A_n,0)	&	(2A_n^\dagger,0,0)	 & -1/2	& -1/4 	& -\alsb\tau/4 \\
	 (2 \de_n,0,0)	&	(2\de_n,0\de_n,0)	  &  1	& 1 		& \alsb(4/\ep+4\Ln+7) \\
 	 (1e_n,1e_n,0)	&	(2e,0_{\bf 8},0_{\bf 8})& -1/2	& -1/2 	& \alsb\tau(-1/\ep-\Ln-2)\\
	 (1f_n, 1f_n,0)	&	(2f,0_{\bf 8},0_{\bf 8} &-1/2	& -1/2 	& \alsb\tau(-1/\ep-\Ln-2)\\
\hline 
	 (1a_\nb,1b_\nb,0)&	 (0,2ab_n,0) 		& -1  & +1 		& \alsb\tau/2\\
	 (1b_\nb,1a_\nb,0) & (0,2ab_\nb^\dagger,0) & -1 & +1 		& \alsb\tau/2\\
	 (2a_\nb,0,0)	&	 (0,2a_\nb,0)	 	 &  1	& -1/2		& \alsb\tau(-1/\ep-\Ln-9/4) \\
	 (0,2a_\nb,0)	&	(0,2a_\nb^\dagger,0)  &  1	& -1/2 	& \alsb\tau(-1/\ep-\Ln-9/4) \\
	 (2A_\nb,0,0)	&	(0,2A_\nb,0)	  	& -1/2	& -1/4 	& -\alsb\tau/4 \\
	 (0,2A_\nb,0)	&	(0,2A_\nb^\dagger,0)  & -1/2	& -1/4 	& -\alsb\tau/4 \\
 	 (2 \de_n,0,0)	&	(2\de_n,0\de_n,0)	  & 1 	& 1		&\alsb(4/\ep+4\Ln+7) \\
\hline
	 (1\de_s,1c_{ns},0)	&	(0c_{ns}\de_s,0,2c_{ns}\de_s)& 1/2 & 1 	& \alsb\tau(-2/\ep-2\Ls-2) \\
	 (2\de c_{sn},0,0)	&	(0c_{ns}\de_s,0,2\de c_{ns}) &  1/ 2 & 1 	& \alsb\tau(-2/\ep-2\Ls-2) \\
	 (1d_{ns}, 1d_{ns},0)	&	0_{\bf 8},0,2d_{ns})	  &  1	&  1	& \alsb\tau(1/\ep+\Ls+1)\\
	 (1d_{\nb s}, 1d_{\nb s},0)	&	(0_{\bf 8},0,2d_{\nb s})  &  1	& 1 	&\alsb\tau(1/\ep+\Ls+1)\\
	 (2A_{ns},0,0)	&	(0,0,2A_{ns})	  & 1	& 1/2					& \alsb\tau(1/\ep+\Ls+1)\\
	 (0,2A_{ns},0)	&	(0,0,2A_{ns}^\dagger)	  & 1	& 1/2			& \alsb\tau(1/\ep+\Ls+1)\\
	 (2A_{\nb s},0,0)	&	(0,0,2A_{\nb s})	  & 1	& 1/2			& \alsb\tau(1/\ep+\Ls+1)\\
	 (0,2A_{\nb s},0)	&	(0,0,2A_{\nb s}^\dagger)	  & 1	& 1/2		& \alsb\tau(1/\ep+\Ls+1)\\
	 (2\de_{s-},0,0)	&	(0,0\de_n,2\de_{s-}) 	  &1 	&1 	&\alsb\tau(-4/\ep-4\Ls-4)\\
	 (2\de_{s\perp},0,0)	&	(-2\de_s,0,4\de_{s\perp})	  & 1/2	& 1/2 	& 4\alsb\tau\\
\hline
	 (1\de_s,0,1n)		&	(-2\de_sM_n,0_{M1},4\de_sM_n)	&1	&1	& 4\alsb\tau\\
	 (0,0,2n_b) 	&	(-2M_{2n_b},0_{M2},4M_{n_b})	&1	&1	&-2\alsb\tau\\
	 (0,0,2\nb_b) 	&	(0,-2M_{\nb_b},4M_{n_b})	&1	&1	&2\alsb\tau\\
	 (0,0,2s_1)	&	(0,0_M,2M_{s1})	&1	& 1 &-4\alsb\tau \\
	 (0,0,2s_{+t})	&	(0,0_M,2M_{+t})	&1	&1 &-4\alsb\tau\\
\hline
\end{array}
}\]\\
\vspace{1\baselineskip}
\[{\renewcommand{\arraystretch}{1.25}
\begin{array}{c || c || c || c}
 (i,j,k) & C_0^{(i,j,k)} & H_0^{(i,j,k)} & R(\tau) \\
 \hline\hline
 (2\de_n,0,0) 	& -4\alsb	&-4\alsb 	& -4\alsb\tau \\
 (1e_n,1e_n,0)	& \alsb	& \alsb 	& \alsb\tau \\
 (1f_n,1f_n,0)	& \alsb	&\alsb	& \alsb\tau \\
 \hline
 (2\de_\nb,0,0)	&-4\alsb	&-4\alsb	& -4\alsb\tau\\
 \hline
 (1d_{ns},1d_{ns},0)	& -\alsb	&-\alsb	&-\alsb\tau\\
 (1d_{\nb s},1d_{\nb s},0)	& -\alsb	&-\alsb	&-\alsb\tau\\
 (2\de_{s-},0,0)	& 4\alsb	&4\alsb	& 4\alsb\tau \\
 \hline
\end{array}
}\]
	\fcaption{\label{tab:match}  The operators and matching coefficients for the $O(\al_s)$ factorization.  The dagger means the operator is the hermitian conjugate.  At the order we are concerned with, there is at most one set of subleading jet and soft operators to be matched onto, so $\al=1$.  The table at the bottom gives the matching coefficients for the last line of \eqn{ope}.  Values of the hard functions, which are defined as $H_{\al,0}^{(i,j,k)}=C_2^{(i)*}C_2^{(j)}C_{\al,0}^{(i,j,k)}$, are also given, as well as the contribution of these operators to $R(\tau)$.}
\end{table*}

The operators in \app{operators} are matched onto combinations of the operators in this section.  The matching coefficients are found by taking the vacuum expectation value of both sides of \eqn{ope} and are shown in \tab{match}.  Using this Table and \tab{vev} the $O(\al_s\tau)$ rate in \eqn{Rnlo} can be calculated.

\ewtxt
\end{appendix}

\bibliography{bibliography}

\end{document}